\documentclass[12pt]{article}

\usepackage{graphics,graphicx,fullpage,natbib,multirow}
\usepackage{amsmath,amssymb,verbatim,epsfig}
\usepackage[dvipsnames,usenames]{color}

\newtheorem{proposition}{Proposition}

\newtheorem{defin}{\bf Definition}

\newenvironment{proof}{\noindent{\bf Proof}}{$\diamond$}

\def\ga{\mbox{Ga}}

\def\no{\mbox{N}}

\def\un{\mbox{Un}}

\def\E{\mbox{E}}
\def\V{\mbox{Var}}

\def\Cr{\mbox{Corr}}

\def\rest{\mbox{rest}}
\def\data{\mbox{data}}
\def\bc{{\bf c}}

\def\bx{{\bf x}}
\def\by{{\bf y}}

\def\bC{{\bf C}}

\def\bM{{\bf M}}

\def\bX{{\bf X}}

\def\bzero{{\bf 0}}
\def\bone{{\bf 1}}
\def\simind{\stackrel{\mbox{\scriptsize{ind}}}{\sim}}

\newcommand{\bbeta}{\boldsymbol{\beta}}
\newcommand{\bgamma}{\boldsymbol{\gamma}}

\newcommand{\Ree}{{\rm I}\!{\rm R}}

\newcommand{\STN}{\mbox{STN}}

\begin{document}

\baselineskip=24pt

\title{\bf Bayesian regression with spatio-temporal varying coefficients}
\author{{\sc Luis E. Nieto-Barajas} \\[2mm]
{\sl Department of Statistics, ITAM, Mexico} \\[2mm]
{\small {\tt lnieto@itam.mx}} \\}
\date{}
\maketitle

\begin{abstract}
To study the impact of climate variables on morbidity of some diseases in Mexico, we propose a spatio-temporal varying coefficients regression model. For that we introduce a new spatio-temporal dependent process prior, in a Bayesian context, with identically distributed normal marginal distributions and joint multivariate normal distribution. We study its properties and characterise the dependence induced. Our results show that the effect of climate variables, on the incidence of specific diseases, is not constant across space and time and our proposed model is able to capture and quantify those changes. 
\end{abstract}

\vspace{0.2in} \noindent {\sl Keywords}: Autoregressive processes,  climate analysis, disease mapping, latent variables, stationary processes.

\section{Introduction}
\label{sec:intro}

The Mexican National Institute for Ecology and Climate Change (INECC) wants to study the impact of climate variables in the health sector in Mexico. In particular, they want to characterise the effect of pluvial precipitation and temperature, on the morbidity of gastrointestinal and respiratory diseases. The study is for each of the 32 states of Mexico and for each month in a window of five years. Previous studies that relate climate variables to diseases are those of \cite{morral&al:18} and \cite{damato&al:14} for gastrointestinal and respiratory diseases, respectively. 

Such a study lies with in the scope of disease mapping methods \citep{lawson:09}. These methods aim to describe and quantify spatial, and sometimes temporal, variation in disease risk, including the identification of patterns of possible association between neighbours in a certain region. As a result of a disease mapping study, it is possible to determine areas, and times, of high and low risk in order to contribute to the disease aetiology. 

Most disease mapping models usually belong to the family of generalized linear mixed model, being the Poisson likelihood the most popular choice. If we denote by $\eta_{i,t}$ an appropriate transformation of the disease (mortality or morbidity) rate for area or location $i$ at time $t$, then the typical specification has the form \citep[e.g.][]{torabi&rosychuk:10} $\eta_{i,t}=\bbeta'\bx_{i,t}+\theta_{i,t}$, where the first part corresponds to the fixed effects, driven by covariates $\bx_{i,t}$, and the second to the random effects. The random effects are, in turn, expressed in terms of spatial effects, temporal effects and sometimes interaction (spatio-temporal) effects. To be specific $\theta_{i,t}=\alpha_i+\delta_t+\xi_{i,t}$. Spatial dependencies $\{\alpha_i\}$ are routinely incorporated into the covariance structure  through normal conditionally auto-regressive (CAR) specifications \citep{besag&al:91,banerjee&al:03}. Temporal dependencies $\{\delta_t\}$ are captured by first order autoregressive (AR), or dynamic, normal models \citep{waller&al:97}. For the interaction effect, proposals are  $\xi_{i,t}=\xi_i t$ or $\xi_{i,t}=\xi_i(t)$, where $\xi_i$ is a spatial (CAR) model and $\xi_i(t)$ is a temporal spline function for each area $i$ \citep[e.g.][]{macnab&dean:01}. 

Random effects are also assumed to be contaminated by observational noise \citep{best&al:05}, or to include seasonal effects \citep{torabi&rosychuk:10}. Multivariate CAR (MCAR) models for the random effects have also been proposed in survival data \citep{jin&carlin:05}. Posterior inference of disease mapping models under a Bayesian approach were optimised using integrated nested Laplace approximations \citep{schrodle&held:10}. Additionally, \cite{goicoa&al:18} studied identifiability constraints in these kinds of models. 

Instead of placing temporal and spatial associations into random effects, they can also be incorporated into the regression coefficients. In other words, the linear predictor would be $\eta_{i,t}=\bbeta_{i,t}'\bx_{i,t}$, where $\bbeta_{i,t}$ is a set of spatio-temporal varying coefficients. In a continuous setting, \cite{gelfand&al:03} used Gaussian processes to define spatial and spatio-temporal varying coefficients with correlations defined via the Mat\`ern function of the spatial and temporal distances. \cite{gelfand&vounatsou:03} defined multivariate conditional autoregressive processes to cope with spatial association and dependence across multiple regression coefficients. \cite{choi&al:12} assumed spatial clusters in which each has a set of homogeneous time-varying coefficients. In contrast, \cite{cai&al:13} generalised parametric to non parametric spatial specifications by considering an area-specific Dirichlet process prior. Under a Bayesian framework, the spatial or spatio-temporal formulations provide easy borrowing of information across the whole study region and across time to provide efficient smooth estimates of the overall spatio-temporal risk patterns as well as variance reduction through the use of shrinkage estimators. 

In this work we introduce a novel spatio-temporal dependent process through a hierarchical model that relies on latent variables. The process has a joint multivariate normal distributions with very flexible associations that can accommodate spatial, temporal and spatio-temporal interaction dependences. We then use it as a prior distribution for spatio-temporal varying coefficients $\bbeta_{i,t}$ in a generalised linear regression framework. We use the proposed model to identify space and time variations in our motivating study of the impact of climate variables in the morbidity of gastrointestinal and respiratory diseases. 

The contents of the rest of the paper is as follows: In Section \ref{sec:prior} we introduce our space-time dependent process and study its association properties. In Section \ref{sec:model} we define the  framework of generalised linear regression models with time-varying coefficients and obtain the corresponding posterior distributions when we use our space-time process as prior distribution. Section \ref{sec:data} contains a detail study of the climate variables impact on the morbidity of some diseases in Mexico. A comparison with alternative priors is also included. We conclude with some remarks in Section \ref{sec:conclusion}. 

Before proceeding we introduce notation. $\no(m,c)$ denotes a univariate normal density with mean $m$ and precision $c$\,; $\no_p(\bM,\bC)$ denotes a $p$-variate normal density with mean vector $\bM_{p\times 1}$ and precision matrix $\bC_{p\times p}$; $\ga(a,b)$ denotes a gamma density with mean $a/b$. The density evaluated at a specific point $x$, will be denoted, for instance for the normal case, as $\no(x\mid m,c)$.

\section{A novel space-time process}
\label{sec:prior}

Let $\beta_{i,t}$ be a parameter of interest for area $i$ at time $t$, for $i=1,\ldots,n$ and $t=1,2,\ldots,T$. The idea is to define a dependence structure among the $\bbeta=\{\beta_{i,t}\}$ in space and time. Since we further want to use this construction as prior distributions for regression coefficients, we also want $\beta_{i,t}$ to have a normal marginal distribution. To achieve this we follow ideas from \cite{jara&al:13}, \cite{nieto&bandyopadhyay:13} and \cite{nieto&huerta:17}.

Let $\partial_{i,t}$ be the set of ``neighbours'', in the broad sense, of area $i$ at time $t$. For spatial dependence we consider actual neighbours, i.e. areas that share a border, plus the current index $i$, whereas for temporal dependence we consider lagged times up to time $t$. Combinations of spatial and temporal dependences are also possible. To be specific, 
\begin{itemize}
\item[$(s)$] Spatial neighbours of areas at the same time, that is \\ $\partial_{i,t}^{(s)}=\{(j,t):\{j\sim i\}\,\cup \{j=i\}\}$, where  ``$\sim$'' denotes spatial neighbour; 
\item[($t_q$)] Temporal neighbours of order $q>0$, that is \\ $\partial_{i,t}^{(t_q)}=\{(i,s):s\in\{t-q,\ldots,t-1,t\}\}$; 
\item[($s+t_q$)] Spatial plus order $q$ temporal neighbours, that is \\ $\partial_{i,t}^{(s+t_q)}=\partial_{i,t}^{(s)}\cup\partial_{i,t}^{(t_q)}$; and 
\item[($s\times t_q$)] Interaction between spatial and temporal dependence or order $q$, i.e., the spatial neighbours are neighbours for all lagged times, that is \\ $\partial_{i,t}^{(s\times t_q)}=\{(j,s):\{j\sim i\}\,\cup \{j=i\};s\in\{t-q,\ldots,t-1,t\}\}$. 
\end{itemize}

Alternative definitions of the set $\partial_{i,t}$ are also possible, for instance seasonal or periodic temporal dependencies \cite[e.g.][]{jara&al:13} or second order spatial dependencies. 
Note that spatial neighbours are reciprocal, that is, $i\sim j$ iff $j\sim i$, whereas temporal neighbours are directed, that is, $s\rightarrow t$ does not imply that $t\rightarrow s$, because we may expect a variable at time $t$ to depend on past values (lagged times $s$), but not the other way around.  

For each area $i$ at time $t$ we require a latent parameter $\gamma_{i,t}$, plus a common parameter $\omega$. Let $\bgamma=\{\gamma_{i,t}\}$, then the proposed model for $\bbeta=\{\beta_{i,t}\}$ is defined through a three level hierarchical model of the form
\begin{eqnarray}
\nonumber
\beta_{i,t}\mid\bgamma&\stackrel{ind}{\sim}&\no\left(\frac{c_0m_0+\sum_{(j,s)\in\partial_{i,t}}c_{j,s}\gamma_{j,s}}{c_0+\sum_{(j,s)\in\partial_{i,t}}c_{j,s}},\,c_0+\sum_{(j,s)\in\partial_{i,t}}c_{j,s}\right)\\
\label{eq:stnormal1}
\gamma_{i,t}\mid\omega&\stackrel{ind}{\sim}&\no(\omega,c_{i,t}) \\
\nonumber
\omega&\sim&\no(m_0,c_0)
\end{eqnarray}
where $m_0\in\Ree$, $c_0>0$ and $c_{i,t}>0$ for $i=1,\ldots,n$ and $t=1,2,\ldots,T$ are known hyper-parameters, and $\partial_{i,t}$ is a set of neighbours which could be any of those defined above. We denote construction \eqref{eq:stnormal1} as $\STN(m_0,c_0,\bc)$, where $\bc=\{c_{i,t}\}$ parameters determine the importance of location $(i,t)$ in the definition of the net, and thus the degree of dependence among $\{\beta_{i,t}\}$. Properties of this construction are given in the following proposition. 

\begin{proposition}
\label{prop:marcor1}
Let $\bbeta\sim\STN(m_0,c_0,\bc)$, that is $\bbeta=\{\beta_{i,t}\}$ for $i=1,\ldots,n$ and $t=1,2,\ldots,T$ is a set of parameters whose joint distribution is defined by \eqref{eq:stnormal1}. Then, $\beta_{i,t}\sim\no(m_0,c_0)$ marginally for all $i$ and $t$. Moreover,  $\bbeta\sim\no_{nT}(\bM_0,\bC)$, that is, $\bbeta$ has a multivariate normal distribution of dimension $nT$ with mean vector $\bM_0=m_0\bone_{nT}$, and $\bone$ a vector of 1's, and covariance matrix $\bC^{-1}$ of dimension $nT\times nT$ with diagonal elements $1/c_0$ and off-diagonal elements appropriately defined by correlations between any two $\beta_{i,t}$ and $\beta_{j,s}$ given by
$$\Cr(\beta_{i,t},\beta_{j,s})=\frac{c_0\left(\sum_{(k,r)\in\partial_{i,t}\cap\partial_{j,s}}c_{k,r}\right)+ \left(\sum_{(k,r)\in\partial_{i,t}}c_{k,r}\right)\left(\sum_{(k,r)\in\partial_{j,s}}c_{k,r}\right)} {\left(c_0+\sum_{(k,r)\in\partial_{i,t}}c_{k,r}\right)\left(c_0+\sum_{(k,r)\in\partial_{j,s}}c_{k,r}\right)},$$
where $\partial_{i,t}$ is a set of neighbours like those defined at the beginning of this Section.
\end{proposition}
\begin{proof}
To prove the marginal distribution we note that, conditionally on $\omega$, dropping the summation indexes, $\sum c_{j,s}\gamma_{j,s}\mid\omega\sim\no\left(\omega\sum c_{j,s}\,,1/(\sum c_{j,s})\right)$ and integrating with respect to $\omega$ we get $\sum c_{j,s}\gamma_{j,s}\sim\no\left(m_0\sum c_{j,s}\,,c_0/\{(\sum c_{j,s})(c_0+\sum c_{j,s})\}\right)$. Now relying on conjugacy properties of the normal model, or simply integrating with respect to $\bgamma$ in the first equation of \eqref{eq:stnormal1}, we obtain the result. For the multivariate normality, we note that $\gamma_{i,t}$ given $\omega$ are conditionally independent, then jointly $\bgamma$, after marginalizing $\omega$, are multivariate normal with common covariance $1/c_0$. Finally, using the same argument, $\beta_{i,t}$'s are conditionally independent given $\bgamma$, after marginalizing $\bgamma$ we obtain that $\bbeta$ are jointly multivariate normal. To obtain the correlation, we first obtain the covariance using conditional expectation properties and then standardise it using the marginal variance. 
\end{proof}

The correlation expression in Proposition \ref{prop:marcor1} has a nice interpretation, the first term in the numerator is a function of the common parameters $c_{k,r}$ that appear in the definition of both $\beta_{i,t}$ and $\beta_{j,s}$ and the second is a function of all dependence parameters in each $\beta_{i,t}$ and $\beta_{j,s}$. In other words, two locations that share the same neighbours will have a higher correlation, even though they are not direct neighbours, and two locations that do not share any neighbour will still be correlated. Additionally, a larger value of $c_{i,t}$ for location $(i,t)$ will make the correlation larger for all pairs of locations that share the same parameter $c_{i,t}$, therefore $c_{i,t}$ can also be seen as a measure of the importance of location $(i,t)$ in the whole network. 

Let us consider a single location so we can concentrate on temporal associations. Let $\bbeta=\{\beta_t\}$ for times $t=1,\ldots,T$ and assume $q=1$. From Proposition \ref{prop:marcor1}, our $\STN$ model implies $\V(\beta_t)=1/c_0$ for all $t$, and correlations for time $t=2$, $\Cr(\beta_2,\beta_3)=\frac{c_0c_2+(c_1+c_2)(c_2+c_3)}{(c_0+c_1+c_2)(c_0+c_2+c_3)}$ and $\Cr(\beta_2,\beta_s)=\frac{(c_1+c_2)(c_{s-1}+c_{s})}{(c_0+c_1+c_2)(c_0+c_{s-1}+c_{s})}$ for $s\geq 4$. If all the $c_{t}$'s had the same value, the correlation between $\beta_2$ and $\beta_3$ would be higher than that between $\beta_2$ and any $\beta_s$ for $s\geq 4$. In general, the correlation can remain high between times $t$ and $s$ even if $|t-s|>q$. This is in contrast to temporal AR models where the correlation decays exponentially to zero as $|t-s|$ increases. 

Now, if we only consider a single time, we can concentrate in spatial associations. Let $\bbeta=\{\beta_i\}$ for areas $i=1,\ldots,n$, then  $\STN$ model \eqref{eq:stnormal1} can be considered as a flexible alternative to the conditional autoregressive (CAR) model. To see this we recall the definition of a CAR model. This is defined through a multivariate normal distribution with mean vector zero and precision matrix $\tau(D_W-\rho W)$, where $W=(w_{ij})$ is the spatial neighbourhood matrix such that $w_{ij}=I(i\sim j)$ and $D_W=\mbox{diag}(w_{1+},\ldots,w_{n+})$ with $w_{i+}=\sum_{j=1}^n w_{ij}$. To better illustrate the differences between $\STN$ and CAR models, let us consider a toy example that consists of a country with only three areas $n=3$, with a neighbourhood structure given by  
\begin{equation}
\label{eq:toy}
\boxed{\beta_1\mid\beta_2\mid\beta_3}.
\end{equation}
From Proposition \ref{prop:marcor1}, $\STN$ model implies a common variance, $\V(\beta_i)=1/c_0$ for $i=1,2,3$, and correlations given by $\Cr(\beta_1,\beta_2)=\frac{c_0(c_1+c_2)+(c_1+c_2)(c_1+c_2+c_3)}{(c_0+c_1+c_2)(c_0+c_1+c_2+c_3)}$, $\Cr(\beta_1,\beta_3)=\frac{c_0c_2+(c_1+c_2)(c_2+c_3)}{(c_0+c_1+c_2)(c_0+c_2+c_3)}$ and $\Cr(\beta_2,\beta_3)=\frac{c_0(c_2+c_3)+(c_1+c_2+c_3)(c_2+c_3)}{(c_0+c_1+c_2+c_3)(c_0+c_2+c_3)}$. On the other hand, the CAR model implies different variances, $\V(\beta_1)=\V(\beta_3)=\frac{2-\rho^2}{2\tau(1-\rho^2)}$ and $\V(\beta_2)=\frac{1}{2\tau(1-\rho^2)}$, and correlations given by $\Cr(\beta_1,\beta_2)=\Cr(\beta_2,\beta_3)=\frac{\rho}{\sqrt{2-\rho^2}}$ and $\Cr(\beta_1,\beta_3)=\frac{\rho^2}{2-\rho^2}$. In our proposed model, spatial dependence is controlled by the set of parameters $\{c_0,c_1,c_2,c_3\}$, whereas in the CAR model it is controlled by a single parameter $\rho$. In both cases, the correlation between two non-neighbouring areas remains positive, however in the CAR model the correlation becomes smaller for further apart areas, whereas in our model the correlation could remain relatively high. 

These features become relevant when we use any of these spatial and /or temporal models as prior distribution for space-time varying coefficients. STN borrows more strength from further apart neighbours, whereas CAR and AR models models mainly borrow strength from close neighbours. Additionally, it might not be justifiable to impose a different prior variance for the coefficients, as the CAR model does.

\section{Regression models}
\label{sec:model}

Let $Y_{i,t}$ be a response variable and $\bX_{i,t}'=(X_{i,t,1},\ldots,X_{i,t,p})$ a set of $p$ covariates for area $i$ at time $t$, for $i=1,\ldots,n$ and $t=1,2,\ldots,T$. We consider a generalised linear model framework \citep[e.g.][]{mccullagh&nelder:89}, so that the response variable $Y_{i,t}$, conditionally on the explanatory variables $\bX_{i,t}$, has a density $f(y_{i,t}\mid\bx_{it})$ which is a member of the exponential family. 

We model the conditional expectation in terms of the explanatory variables as $\E(Y_{i,t}\mid\bX_{it})=g^{-1}(\eta_{i,t})$, where $g(\cdot)$ is an appropriate link function and $\eta_{i,t}$ is a linear predictor, which for static coefficients has the form $\eta_{i,t}=\alpha+\bbeta'\bx_{i,t}$. We introduce a space-time dynamic in the regression coefficients and define the linear predictor as 
\begin{equation}
\label{eq:predictor}
\eta_{i,t}=\alpha+\bbeta_{i,t}'\bx_{i,t}. 
\end{equation}

To state the prior for the model parameters we expand the inner product in the linear predictor and write $\eta_{i,t}=\alpha+\beta_{i,t,1}x_{i,t,1}+\cdots+\beta_{i,t,p}x_{i,t,p}$ and take $\alpha\sim\no(m_\alpha,c_\alpha)$ for the intercept and for the regression coefficients of each explanatory variable $k$ we take a normal space-time processes, that is, $\bbeta_k=\{\beta_{i,t,k}\}\sim\STN(m_0,c_0,\bc)$ defined by \eqref{eq:stnormal1}, independently for $k=1,\ldots,p$.  

The likelihood for the regression model is simply $f(\by\mid\bx)=\prod_i\prod_t f(y_{i,t}\mid \bx_{i,t})$, and the joint prior for all parameters, including the latent ones, has the form 
$$\hspace{-7cm}f(\alpha,\bbeta,\bgamma,\omega)=\no(\alpha\mid m_\alpha,c_\alpha)\left\{\prod_k\no\left(\omega_k\mid m_0,c_0\right)\right\}$$ $$\hspace{7mm}\times\left\{\prod_i\prod_t\prod_k\no\left(\beta_{i,t,k}\left|\frac{c_0m_0+\sum_{(j,s)\in\partial_{i,t}}c_{j,s}\gamma_{j,s,k}}{c_0+\sum_{(j,s)\in\partial_{i,t}}c_{j,s}},c_0+\sum_{(j,s)\in\partial_{i,t}}c_{j,s}\right.\right)\no\left(\gamma_{i,t,k}\mid\omega_k,c_{i,t}\right)\right\}.$$
Alternatively, after integrating with respect to the latent parameters $(\bgamma,\omega)$, the prior becomes $f(\alpha,\bbeta)=\no(\alpha\mid m_\alpha,c_\alpha)\prod_{k=1}^p\no_{nT}(\bbeta_k\mid\bM_0,\bC)$.

To specify the posterior distributions induced, let us consider a normal likelihood for the regression model of the form 
\begin{equation}
\label{eq:normal}
Y_{i,t}\mid\bx_{i,t}\sim\no(\mu(\bx_{i,t}),\tau_{i,t}) 
\end{equation}
for $i=1,\ldots,n$ and $t=1,\ldots,T$, where $\mu(\bx_{i,t})=\eta_{i,t}$ is the mean defined by an identity function of the linear predictor \eqref{eq:predictor}, and $\tau_{i,t}$ is the precision parameter. In this case, the conditional posterior distribution for each $\bbeta_k$  becomes a multivariate normal of dimension $nT$. For large $n$ and $T$, which is the case of our motivating example where $nT=32\times60=1920$, dealing with a multivariate normal of such dimension is not computationally feasible. Alternatively, we suggest to sample from the univariate conditional posterior distributions for each $\beta_{i,t,k}$ given the latent parameters and the data. The full conditional distributions for all model parameters are:
\begin{enumerate}
\item[i)] $f(\alpha\mid\by,\rest)=\no(\alpha\mid\mu_\alpha,\tau_\alpha)$, where $$\mu_\alpha=\frac{c_\alpha m_\alpha+\sum_i\sum_t\tau_{i,t}\left(y_{i,t}-\sum_k\beta_{i,t,k}x_{i,t,k}\right)}{c_\alpha+\sum_i\sum_t\tau_{i,t}}\quad\mbox{and}\quad \tau_\alpha=c_\alpha+\sum_i\sum_t\tau_{i,t}$$
\item[ii)] $f(\beta_{i,t,k}\mid\by,\rest)=\no(\beta_{i,t,k}\mid\mu_\beta,\tau_\beta)$, where
$$\mu_\beta=\frac{c_0m_0+\sum_{(j,s)\in\partial_{i,t}}c_{j,s}\gamma_{j,s,k}+\tau_{i,t}\,x_{i,t,k}(y_{i,t}-\alpha-\sum_{l\neq k}\beta_{i,t,l}x_{i,t,l})}{c_0+\sum_{(j,s)\in\partial_{i,t}}c_{j,s}+\tau_{i,t}\,x_{i,t,k}^2}$$ and
$$\tau_\beta=c_0+\sum_{(j,s)\in\partial_{i,t}}c_{j,s}+\tau_{i,t}\,x_{i,t,k}^2$$
for $i=1,\ldots,n$, $t=1,\ldots,T$ and $k=1,\ldots,p$.
\item[iii)] $f(\gamma_{i,t,k}\mid\by,\rest)=\no(\gamma_{i,t,k}\mid\mu_\gamma,\tau_\gamma)$, where 
$$\mu_\gamma=\frac{\omega_k+\sum_{(j,s)\in\varrho_{i,t}}\left\{\beta_{j,s,k}-\frac{1}{D_{j,s}}\left(c_0m_0+\sum_{(l,u)\neq(i,t)\in\partial_{j,s}}c_{l,u}\gamma_{l,u,k}\right)\right\}}{1+c_{i,t}\sum_{(j,s)\in\varrho_{i,t}}\frac{1}{D_{j,s}}}$$
and $$\tau_\gamma=c_{i,t}\left(1+c_{i,t}\sum_{(j,s)\in\varrho_{i,t}}\frac{1}{D_{j,s}}\right),\quad\mbox{with}\quad D_{j,s}=c_0+\sum_{(l,u)\in\partial_{j,s}}c_{l,u},$$
where $\varrho_{i,t}$ is the set of reversed neighbours, that is, the set of pairs $(j,s)$ such that $(i,t)\in\partial_{j,s}$, 
for $i=1,\ldots,n$, $t=1,\ldots,T$ and $k=1,\ldots,p$.
\item[iv)] $f(\omega_k\mid\by,\rest)=\no(\omega_k\mid\mu_\omega,\tau_\omega)$, where
$$\mu_\omega=\frac{c_0m_0+\sum_i\sum_t c_{i,t}\gamma_{i,t,k}}{c_0+\sum_i\sum_t c_{i,t}}\quad\mbox{and}\quad\tau_\omega=c_0+\sum_i\sum_t c_{i,t}$$
for $k=1,\ldots,p$.
\end{enumerate}

Finally, we assume $\tau_{i,t}=\tau_i$ to have a common precision along time for each area $i$. If we further take $\tau_i\sim\ga(a_0,b_0)$ a-priori then its conditional posterior distribution has the form
\begin{enumerate}
\item[v)] $f(\tau_i\mid\by,\rest)=\ga(\tau_i\mid a_\tau,b_\tau)$, where
$$a_\tau=a_0+\frac{T}{2}\quad\mbox{and}\quad b_\tau=b_0+\frac{1}{2}\sum_t\left(y_{i,t}-\alpha-\sum_k\beta_{i,t,k}x_{i,t,k}\right)^2$$
for $i=1,\ldots,n$.
\end{enumerate}

Posterior inference of the model parameters relies on a Gibbs sampler \citep{smith&roberts:93}. Conditional posterior distributions i)--v) are of standard form because we are dealing with a normal regression model for which our normal and space-time normal prior distributions are conditionally conjugate. For other likelihoods different to the normal, say Bernoulli, Poisson or gamma, distributions i) and ii) must be appropriately modified and a Metropolis-Hastings step \citep{tierney:94} would be required to sample from them.

\section{Numerical analyses}
\label{sec:data}

\subsection{Simulation studies}

We carry out two simulation studies to test the performance of our posterior simulation procedure. We consider a simple spatial setting with $n=3$ regions as in \eqref{eq:toy}, plus the temporal dimension. The first scenario consists in fixing the regression coefficients $\beta_{i,t}$ in a deterministic way as follows: $\beta_{1,t}=(1.1)^t$, $\beta_{2,t}=1+(1.01)^t$ and $\beta_{3,t}=(0.9)^t$, for $t=1,\ldots,T$ with $T=10$. We then sample response variables from model \eqref{eq:normal} with $\alpha=-1$ and $\tau_{i,t}=\tau_i=1$ and a single covariate $x_{i,t}\sim\un(0,1)$ to form a sample of size $nT=30$. This experiment was repeated 100 times. 
The prior distribution was defined by $m_\alpha=0$, $c_\alpha=0.01$, $m_0=0$, $c_0=0.01$, $c_{i,t}=0.1$ and $a_0=b_0=0.01$. The sets of neighbours were those defined in Section \ref{sec:model} with $q=1,2$. Posterior inference was obtained through a Gibbs sample with two parallel chains of 21,000 iterations with a burn-in of 1,000 and keeping one of every 10$^{th}$ iteration.

For model comparison we compute two goodness of fit (gof) measures, the logarithm of the pseudo marginal likelihood (LPML) and the deviance information criterion (DIC). LPML is a summary measure of conditional predictive ordinates (CPO), commonly used for model comparison and introduced by \cite{geisser&eddy:79}. Given posterior samples of model parameters, $\alpha^{(r)}$, $\beta_{i,t,k}^{(r)}$ and $\tau_i^{(r)}$ for $r=1,\ldots,R$, a Monte Carlo estimate $\widehat{\mbox{CPO}}_{i,t}$, for each data point $(i,t)$, is obtained as $$\widehat{\mbox{CPO}}_{i,t}=\left(\frac{1}{R}\sum_{r=1}^R\frac{1}{\no\left(y_{it}\mid\alpha^{(r)}+\bbeta_{i,t}^{(r)'}\bx_{i,t},\tau_i^{(r)}\right)}\right)^{-1},$$
for $i=1,\ldots,n$ and $t=1,\ldots,T$. Finally these values are summarised to define $\mbox{LPML}=\sum_{i=1}^n\sum_{t=1}^T\log(\mbox{CPO}_{i,t})$. On the other hand, DIC was introduced by \cite{spiegelhalter&al:02} and is a model selection criterion that penalises for model complexity. Larger/smaller values of LPML/DIC measures are preferable. 

\begin{figure}
\begin{center}
\includegraphics[scale=0.6]{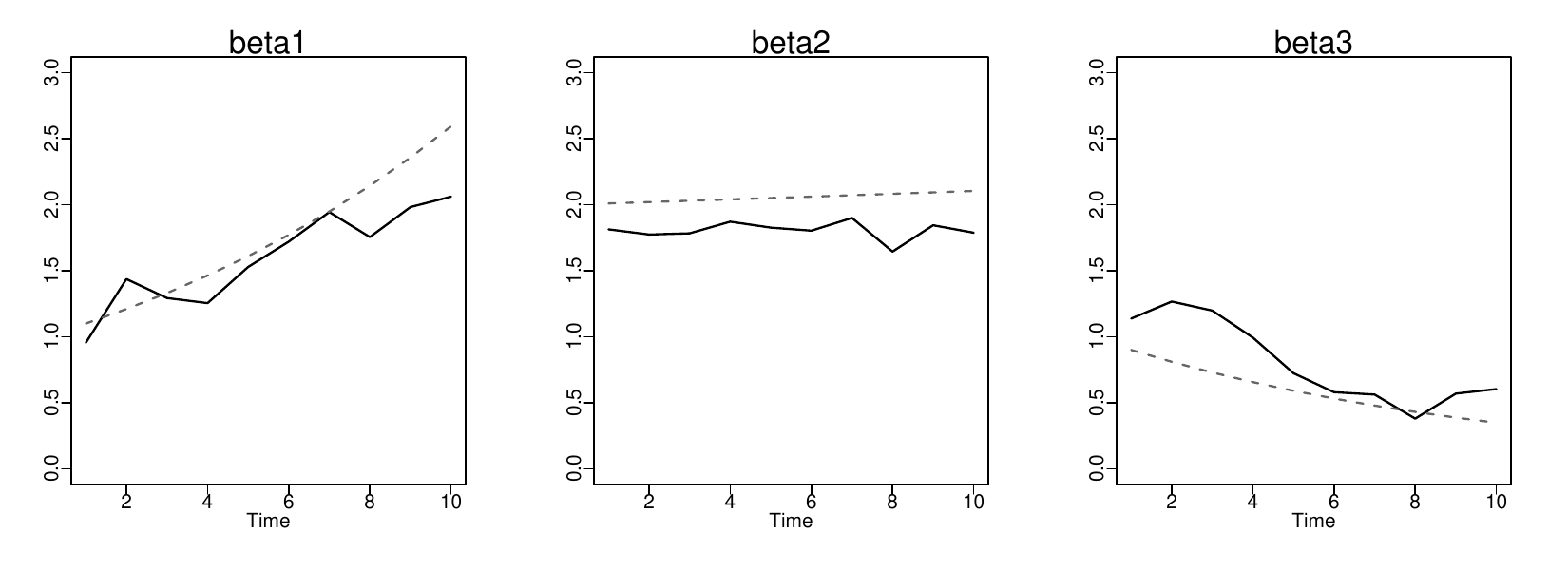}
\caption{\small{Simulated scenario 1. $\beta_{i,t}$ for $i=1,2,3$ and $t=1,\ldots,10$. Posterior means, averaged across the 100 experiments (solid line), and real values (dotted line).}}
\label{fig:betasim}
\end{center}
\end{figure}

Table \ref{tab:sim} reports the average LPML and DIC across the 100 repetitions of the experiment. Looking at the third and fourth columns we observe that the best fitting is obtained with a temporal neighbourhood structure $(t_q)$ with $q=2$. This makes sense since there is no relation among spatial neighbours in the definition of the coefficients. For this wining model, the average coverage of 95\% credible intervals (CI) for all $\beta_{i,t}$ is, in average, 93.3\%. Figure \ref{fig:betasim} includes posterior estimates (averaged across the experiments). Our estimates capture the temporal trends in the three regions. 

The second scenario consists in random coefficients simulated from model \eqref{eq:stnormal1} with parameters $m_0=0$, $c_0=1$, $c_{i,t}=1$ and neighbourhood structure $(s+t_q)$ with $q=1$. The data were generated as in the first scenario and 100 repetitions of the experiment were also obtained. The objective of this study is to see whether the gof measures are able to detect the correct neighbourhood structure. Therefore we took the same specifications in the prior as those used to simulate the coefficients $\beta_{i,t}$, plus $m_\alpha=0$, $c_\alpha=0.01$ for $\alpha$, and $a_0=b_0=0.01$ for $\tau_i$. We tried different sets of neighbours, as those defined in Section \ref{sec:model} with $q=1,2$, and the same Gibbs sampler specifications as above. 

Results are shown in the fifth and sixth columns in Table \ref{tab:sim}. The best model chosen by both the LPML and the DIC is that with neighbourhood $(s+t_q)$ and $q=1$ as it should be.

\subsection{Climate impact on morbidity}

In general, the climate in a region is characterised by the atmospheric conditions of temperature and pluvial precipitation, among other factors. The Mexican Autonomous National University (UNAM) through the group of Climate and Society of the Center for Atmospheric Sciences has created a monthly database of climate records for all 32 states of Mexico from 1901 to the date. Measured variables are average and maximum temperature, both in centigrades degrees, and average pluvial precipitation in millimetres. 

Public health records in Mexico are obtained through the National System for Epidemiology Surveillance (SINAVE). This system gathers information from the whole country and all institutions of the national system of health. Information is available from 1985 to date for all 32 states of Mexico. We will concentrate in two big groups of diseases, gastrointestinal and respiratory. Each of this groups is formed by several specific diseases, say: amebiasis, cholera, typhoid, intestinal infection, intoxication, salmonellosis and shigellosis, for the gastrointestinal group; and asthma, pharyngitis, acute respiratory infections, influenza and pneumonia, for the respiratory group. 

Since the information of all specific diseases is not available for the same years, we will concentrate on a window of five years, from 2011 to 2015 (60 months) for the two disease groups. Additionally, we will require two more variables, the percentage of illiterate people as a marginality indicator, to capture the trend, and the population size (number of inhabitants) as an offset. 

We now apply our Bayesian regression model, with spatio-temporal varying coefficients and STN prior, to the motivating study of climate impact on morbidity of gastrointestinal and respiratory diseases in Mexico, mentioned in Section \ref{sec:intro}. 

To state the model we define variables: $Y_{i,t}=\log\left(N_{i,t}/P_{i,t}\right)$ is the morbidity rate (in log scale), where $N_{i,t}$ is number of disease cases and $P_{i,t}$ is the population size; $X_{i,t,1}$ is the average pluvial precipitation (in log scale); $X_{i,t,2}$ is the average temperature (in log scale); and $X_{i,t,3}$ is the percentage of illiterate people (in log scale), for state $i=1,\ldots,n$ and month $t=1,\ldots,T$, with $n=32$ and $T=60$. The maximum temperature will not be used in the model due to a high correlation with the average temperature. The model is therefore as in equations \eqref{eq:predictor} and \eqref{eq:normal} with $p=3$ and $\tau_{i,t}=\tau_i$. 

\begin{figure}
\begin{center}
\includegraphics[trim=2cm 3cm 0cm 0cm,scale=0.75]{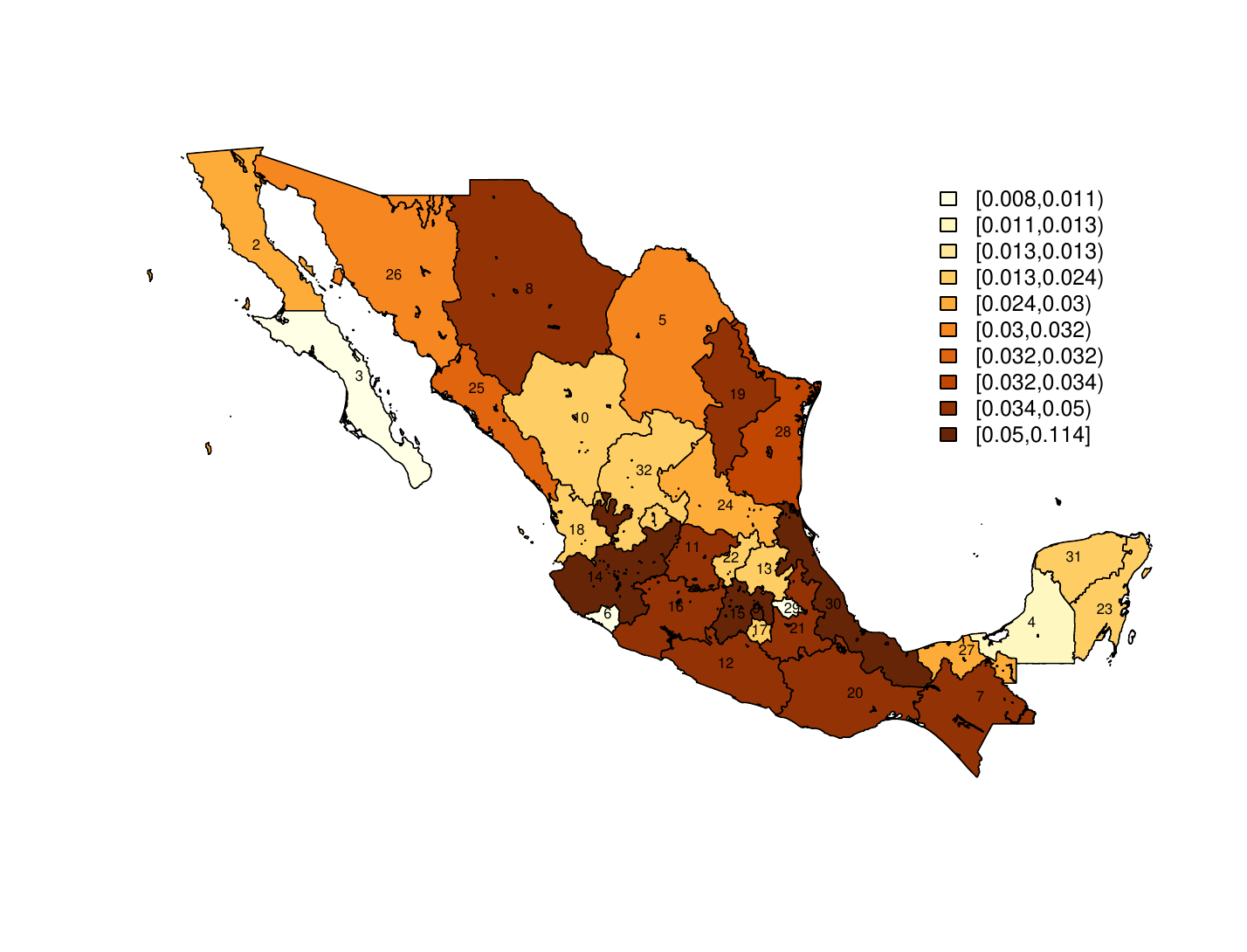}
\caption{\small{Proportion of cases, aggregating all years, for each state. Gastrointestinal group.}}
\label{fig:pcases}
\end{center}
\end{figure}

Considering the 32 states and the 60 months we have around 28 million cases registered in the gastrointestinal group and 133 million cases in the respiratory group. From this total, the percentage of cases for each state is presented in Figure \ref{fig:pcases}. We produced this graph for both, the gastrointestinal and the respiratory groups, however they were almost identical, confirming that the number of cases is a function of the number of people at risk (population size). Therefore we only present the graph for the gastrointestinal group. The state with the largest proportion of cases, is the State of Mexico ($i=15$) with around 11\% of the cases, followed by Mexico City ($i=9$) with around 7\% of the cases. The smallest states, in terms of population size are Baja California Sur ($i=3$) and Colima ($i=6$), which present the smallest proportion with less than 1\% of the cases. 

\begin{figure}
\begin{center}
\includegraphics[scale=0.335]{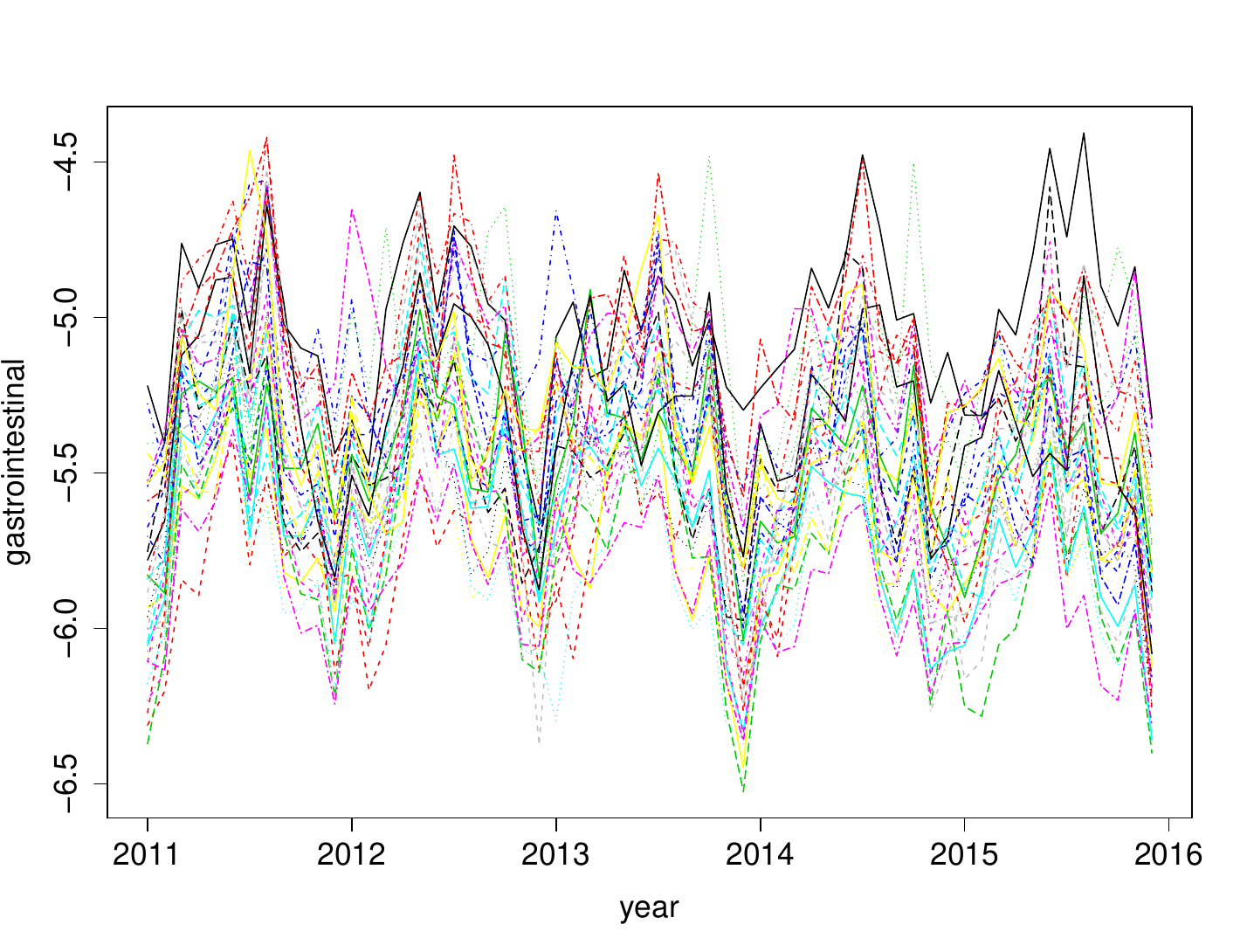}
\includegraphics[scale=0.335]{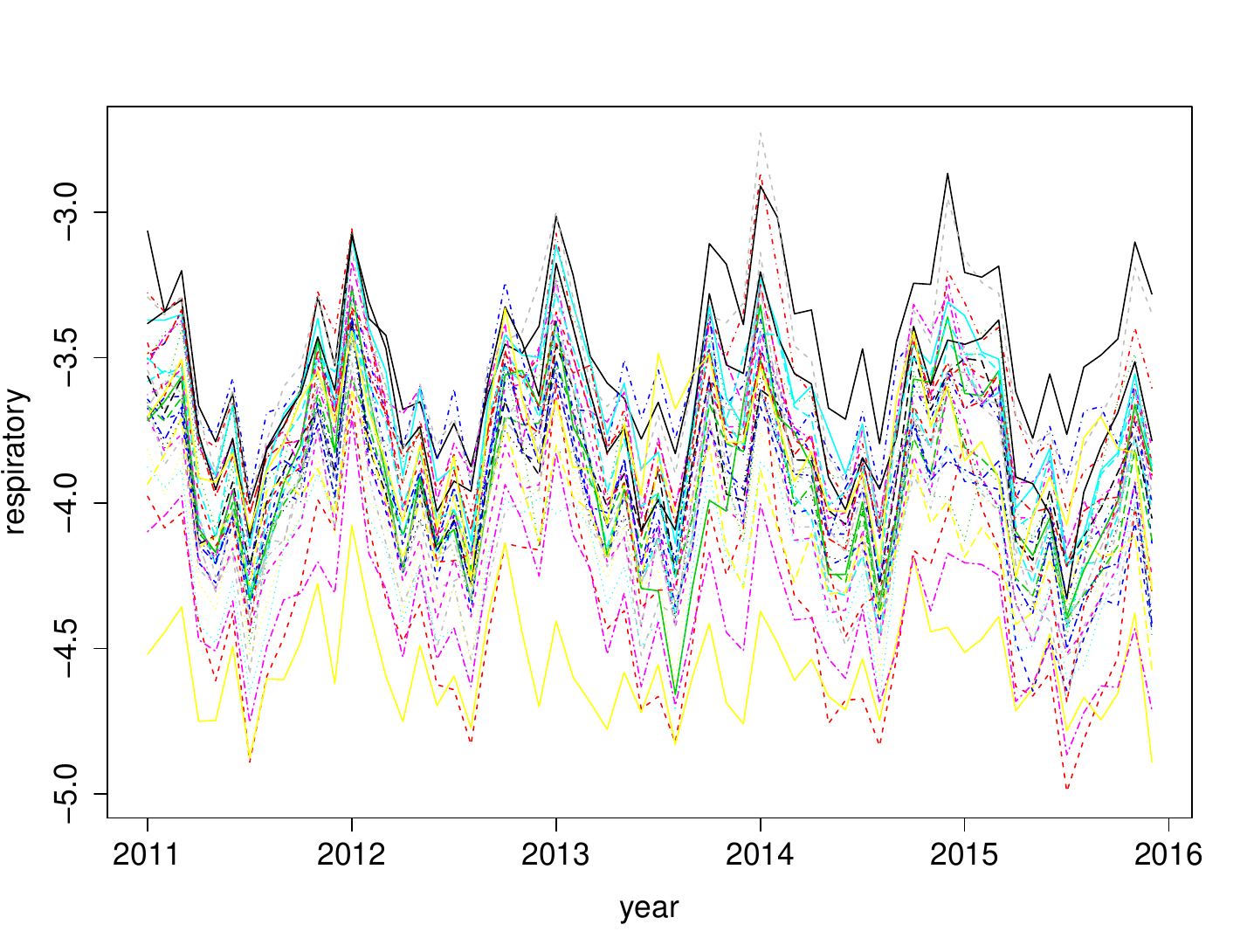}
\includegraphics[scale=0.22]{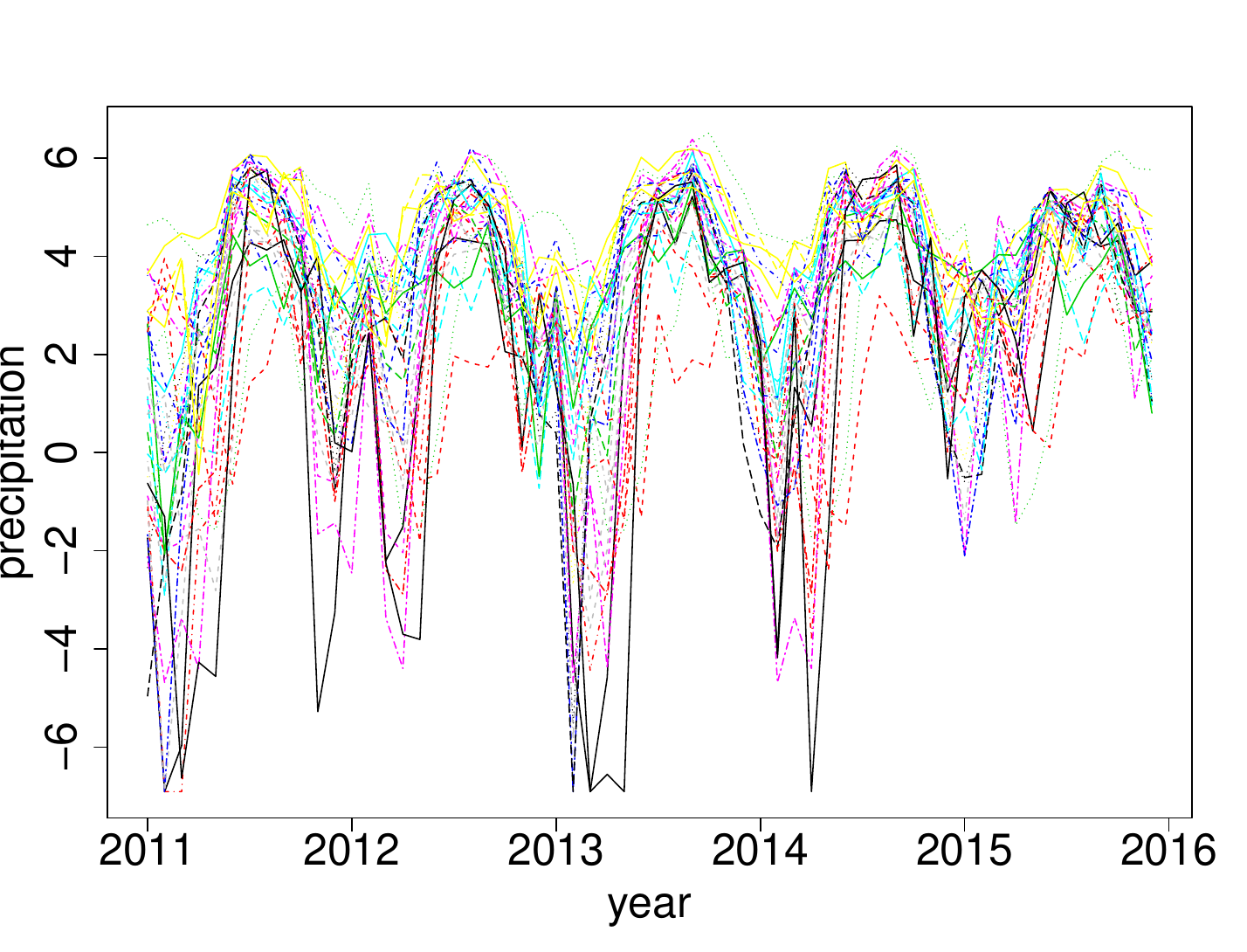}
\includegraphics[scale=0.22]{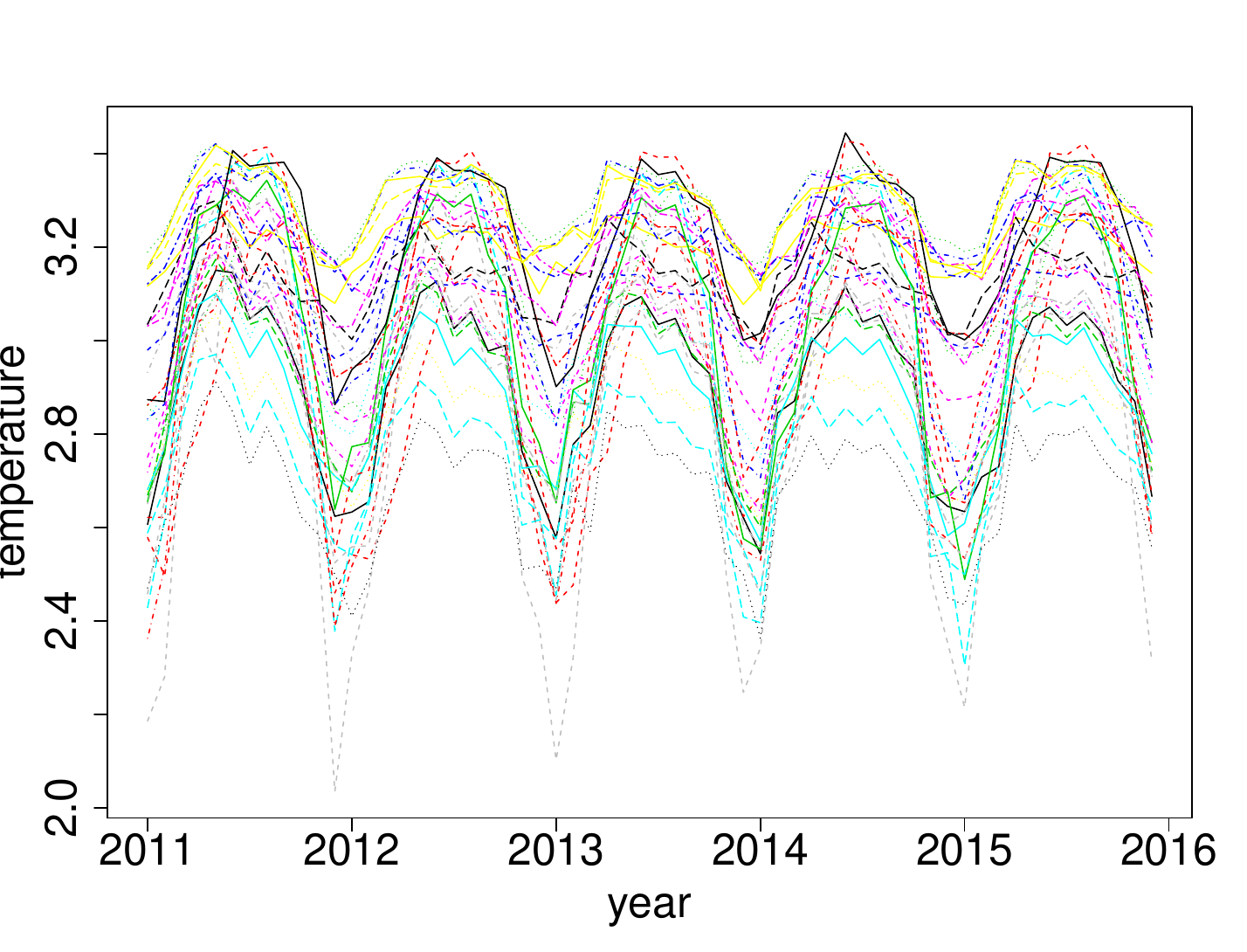}
\includegraphics[scale=0.22]{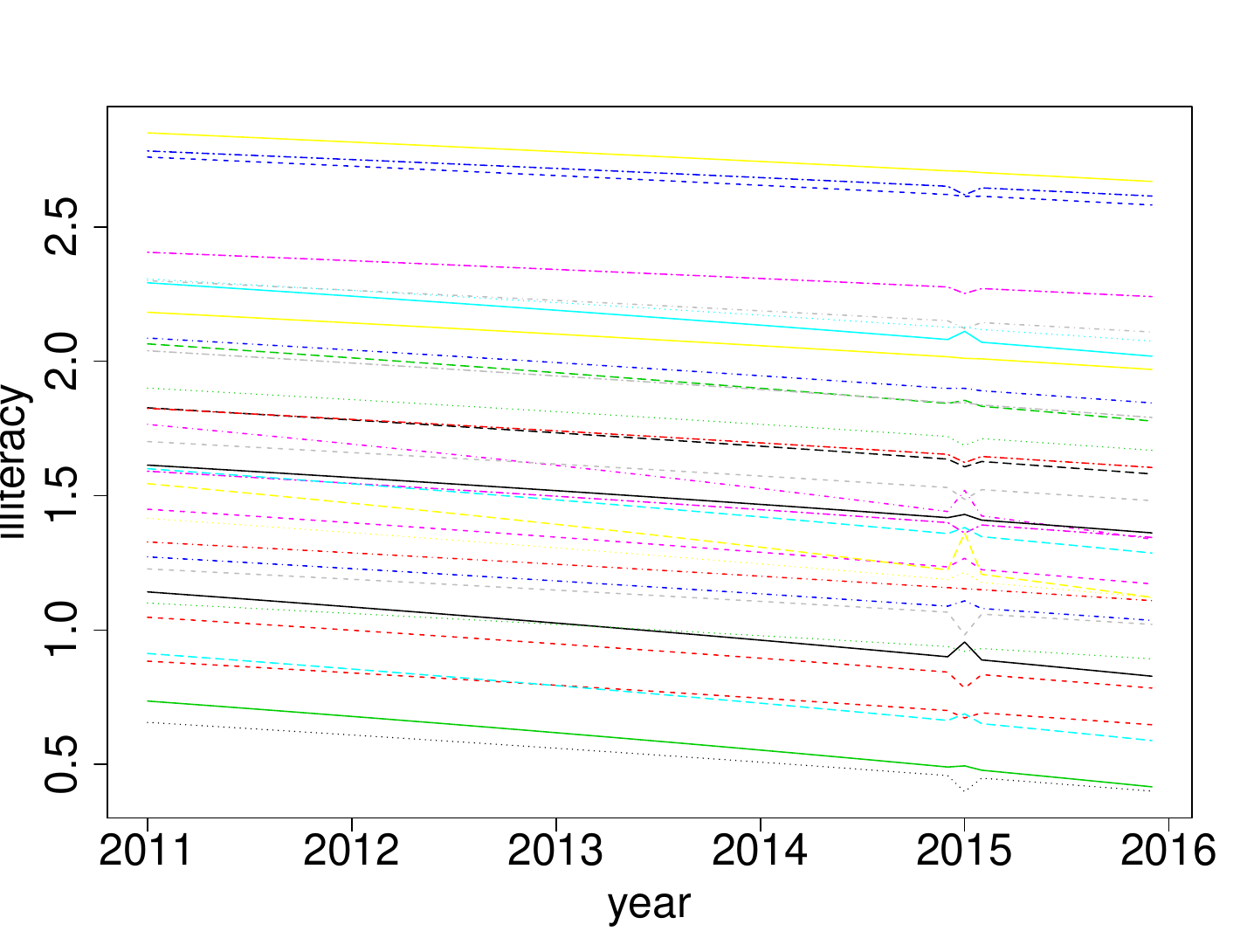}
\caption{\small{Time series of the data. Top row: gastrointestinal rates (left), respiratory rates (right). Bottom row: pluvial precipitation (left), averate temperature (middle), and illiteracy proportion (right). All in log scale.}}
\label{fig:tseries}
\end{center}
\end{figure}

The variables involved in the model are presented in Figure \ref{fig:tseries} as time series. In the panels we show $Y_{i,t}$ for the gastrointestinal and respiratory groups in the top row, and the three explanatory variables $X_{i,t,k}$ for $k=1,2,3$ in the bottom row. Apart from the illiteracy proportion (bottom right panel) which shows a decreasing trend, the rest of the variables present a 12 months seasonal pattern. We also note that the log rates for gastrointestinal and respiratory groups are shifted a period of 6 months. This is explained by the fact that respiratory infections have a peak in the winter, whereas gastrointestinal cases occur more often in the summer. 

To specify the model, we consider the four sets of neighbours defined in Section \ref{sec:prior}, $(s)$, $(t_q)$, $(s+t_q)$ and $(s\times t_q)$. In particular we took temporal neighbours that define a short term dependence, lags up to a quarter year, or a medium term dependence, lags up to a semester, i.e. $q\in\{3,6\}$ to compare; and spatial neighbours given by the geographical adjacencies of the 32 states of Mexico, which are included in Table \ref{tab:adj}. The number of adjacencies per state goes from 1 to 8 with a median value of 4.

Prior distributions for our model parameters were defined by: $m_\alpha=0$ and $c_\alpha=0.01$, for $\alpha$; $m_0=0$, $c_0=0.01$ and constant $c_{i,t}$ $\forall i,t$ with values in $\{10,50,100\}$ to define low, medium and high prior dependence, for $\bbeta_k$; and $a_0=b_0=0.01$, for $\tau_i$. Posterior inference requires the implementation of a Gibbs sampler with a very large number of parameters, therefore we expect to experience high autocorrelation in the chains. We ran two parallel chains for $115,000$ iterations with a burn-in of $15,000$ and kept one of every $50^{th}$ iteration. Convergence of the chains was assessed by monitoring the trace plots, the ergodic means, as well as the autocorrelation function. The code is written in Fortran and is available as Supplementary Material. 

For comparison purposes, we considered three competing priors: 
$\bbeta_{t,k}=\{\beta_{i,t,k},\,i=1,\ldots,n\}\sim\no_n(\bzero,\tau_c(D_W-\rho W))$, i.e. CAR priors to account for spatial dependence; $\bbeta_{i,k}=\{\beta_{i,t,k}\,t=1,\ldots,T\}\sim\no_T(\bzero,\Sigma^{-1})$, i.e. stationary first order AR processes, to account for temporal dependence, defined by covariance elements $\sigma_{t,s}=\varphi^{|t-s|}/\tau_{ar}$; and $\bbeta_k=\{\beta_{i,t,k},\,i=1,\ldots,n,\,t=1,\ldots,T\}\sim\no_{nT}(\bzero,(D_W-\rho W)\otimes\Sigma^{-1})$, i.e. MCAR priors \citep{gelfand&vounatsou:03} to account for space and time dependence, with covariance matrix $\Sigma$ defined as in the first order AR process. These priors were specified by taking $\tau_c=\tau_{ar}=0.1$ and $\rho=\varphi=0.99$. 

Additionally, we also considered two baseline models: Model 0, which assumes a different regression coefficient for each individual $i$ but common for all times $t$. In notation, the linear predictor for this model is $\eta_{i,t}=\alpha+\bbeta_i'\bx_{i,t}$ and the precision is $\tau_i$; and model 00 that assumes a common regression coefficient for all individuals $i$ and for all times $t$. This is obtained with linear predictor $\eta_{i,t}=\alpha+\bbeta'\bx_{i,t}$ and precision $\tau$. For these two models we took $\no(0,0.01)$ independent prior distributions for all parameters $\alpha$, $\beta_{i,k}$ and $\beta_k$, respectively, and $\ga(0.01,0.01)$ priors for the precisions $\tau_i$ and $\tau$, respectively. 

We assess model fit by computing the two gof measures, LPML and DIC. These are reported in Table \ref{tab:gof} for the 21 different versions of our model, the three competitors, plus models 0 and 00, all fitted to both datasets, gastrointestinal and respiratory. Additionally, Table \ref{tab:gof} reports the median size of the neighbourhoods, $\mbox{med}|\partial_{i,t}|$, that define each of our models. For spatial models $(s)$ the neighbourhood size ranges from 2 to 9, so the median size reported is 5. For temporal models $(t_q)$ the exact neighbourhood size is $q+1$. 

In all cases, the two gof measures are in agreement in selecting the best model. For both datasets, the worst model is model 00, however model 0 is better than some versions of our space-time model. In general we note that models with larger dependence parameters $c_{i,t}$ equal to 50 or 100 achieve better fitting in models with small neighbourhood size, which is the case for spatial $(s)$, temporal $(t_q)$, and spatial plus temporal $(s+t_q)$ models. However, for models with space-time interaction $(s\times t_q)$, which have larger neighbourhood size, the smaller value $c_{i,t}=10$ is preferred. In other words, as the number of neighbours increases, both datasets prefer models with smaller values of the dependence parameters. 

Considering the best spatial model $(s)$ with $c_{i,t,}=50$, we obtain a better fit than model 0, for gastrointestinal data, but a worst fit for respiratory data. This tell us that, somehow, gastrointestinal diseases are more spatially dependent than respiratory diseases. On the other hand, the best temporal model $(t_q)$ for gastrointestinal data is obtained with a lag of order $q=3$ and a dependence parameter $c_{i,t}=100$, whereas for respiratory data it is obtained with $q=6$ and $c_{i,t}=50$. In both cases, the best temporal model is better than model 0.

Now, comparing the space and time models $(s+t_q)$ and $(s\times t_q)$, the former achieves better fit. In fact, for both datasets the best model, overall, is obtained with a neighbourhood that considers spatial plus medium term neighbours in time (order $q=6$), and with a medium strength prior dependence ($c_{i,t}=50$). In other words, the best model is choosing a neighbourhood structure for the pair $(i,t)$ that depends on the same state $i$ in the previous 6 months and on the geographical adjacent states for the same time $t$. This model has a median neighbourhood size of 11 and obtains gof statistics of, LPML$=974$ and DIC$=-2199$, for the gastrointestinal data, and LPML$=982$ and DIC$=-2148$, for the respiratory data. 

\begin{figure}
\begin{center}
\includegraphics[scale=0.45]{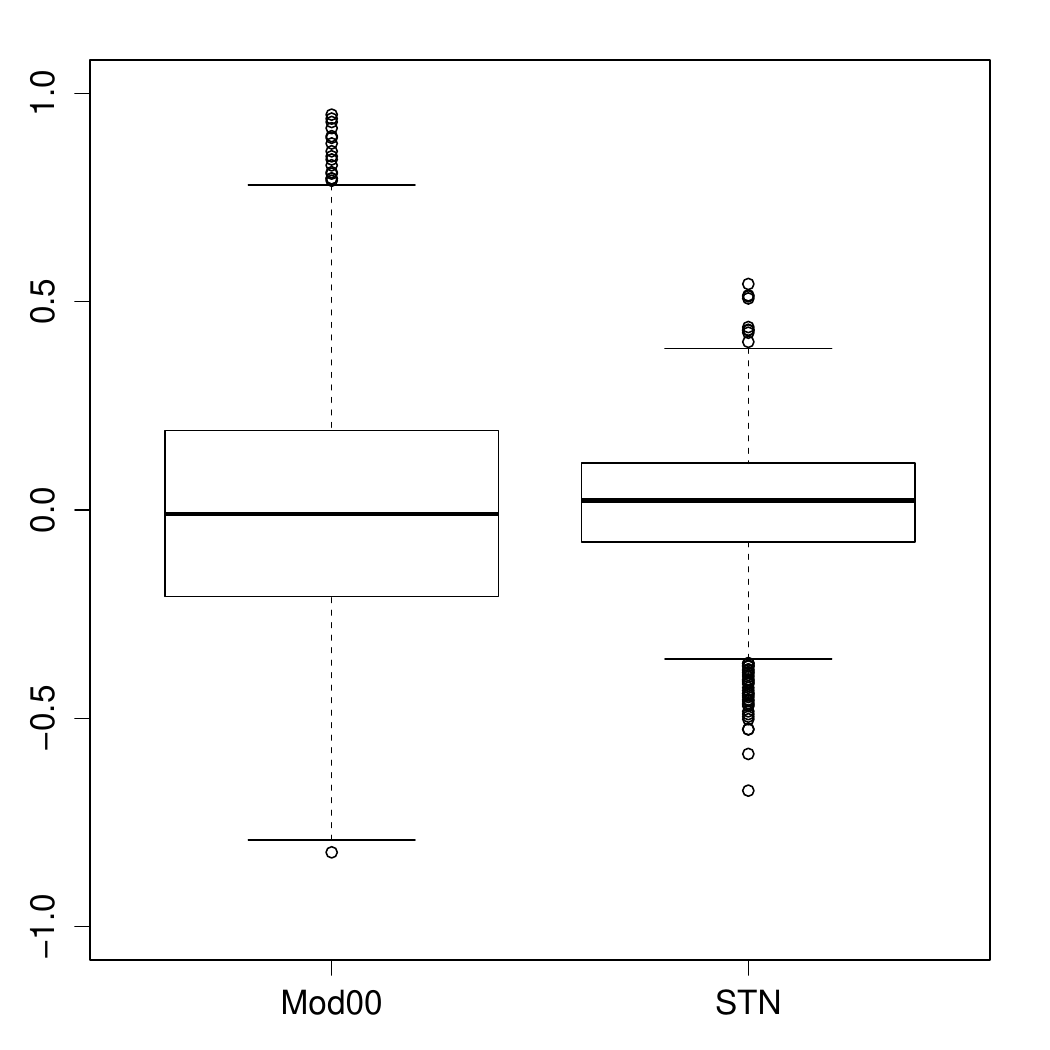}
\includegraphics[scale=0.45]{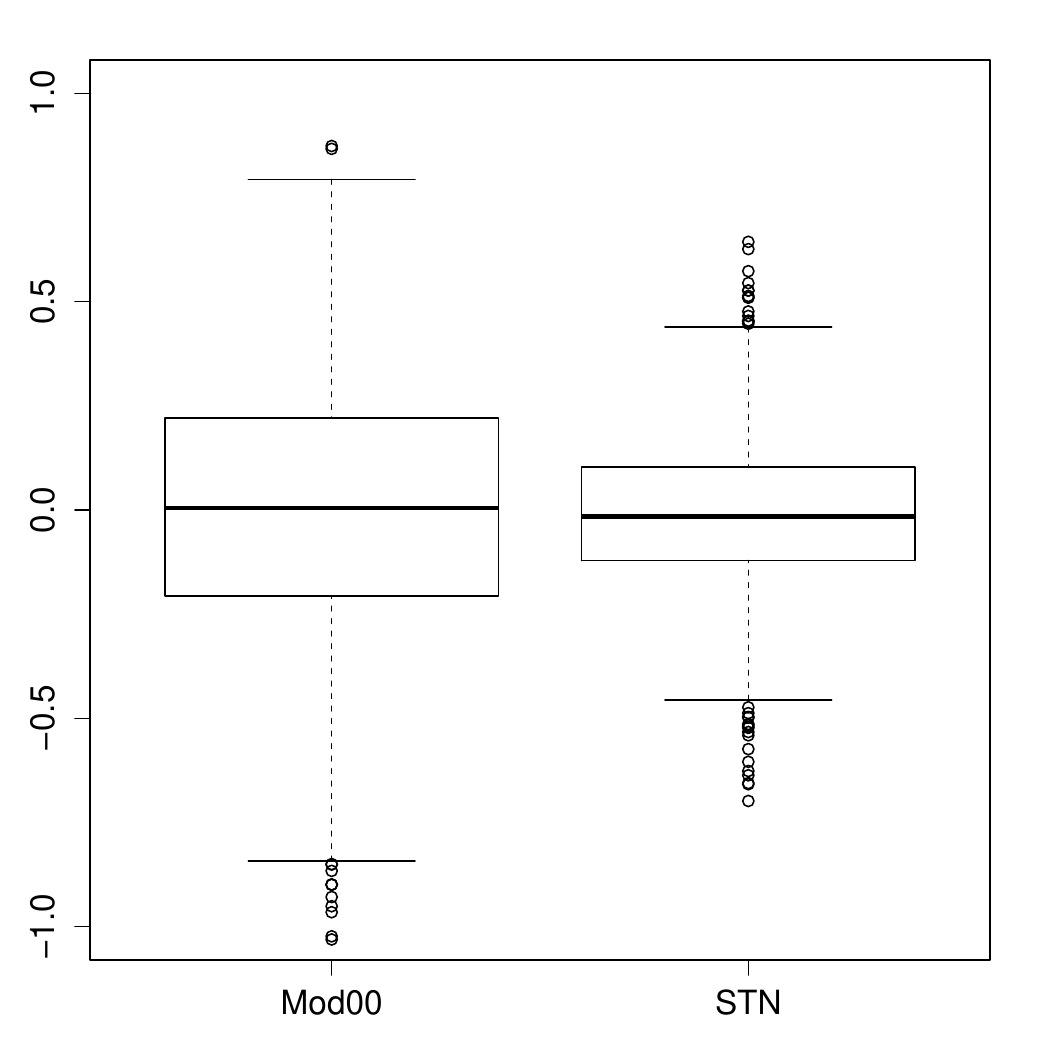}
\caption{\small{Box plot of residuals for model 00 and STN with neighbourhood $(s+t_q)$ and $q=6$. Gastrointestinal data (left) and respiratory data (right).}}
\label{fig:res}
\end{center}
\end{figure}

An alternative way of assessing the gain in the fitting, we computed the residuals $y_{i,t}-E(y_{i,t}\mid\data)$ for model 00 and our best fitting model STN. Box plots of them are presented in Figure \ref{fig:res}. For both datasets the variance not explained by the model is highly reduced when using our STN model. 

Analysing the gof values of the three competing priors, we note that none of them achieve a good fitting. AR prior behaves better than the CAR prior, and both priors combined in the MCAR is the best among these three. The performance of MCAR is compared to that of our model with only spatial dependence with association parameters $c_{i,t}=10$, for both datasets. 

For reference purposes, posterior 95\% credible intervals for the model parameters under model 00 are: $\alpha\in(-7.91,-7.59)$, $\beta_1\in(-0.018,-0.006)$, $\beta_2\in(0.79,0.91)$, $\beta_3\in(-0.18,-0.13)$, for gastrointestinal data; and $\alpha\in(-2.44,-2.04)$, $\beta_1\in(-0.04,-0.03)$, $\beta_2\in(-0.57,-0.44)$, $\beta_3\in(-0.04,0.01)$, for respiratory data. From these numbers we can say that pluvial precipitation ($\bX_{1}$) has a negative effect in both disease groups, whereas temperature ($\bX_{2}$) has a different effect, it is positive for gastrointestinal data and negative for respiratory data, which makes sense. Finally, the percentage of illiterate people ($\bX_{3}$), has a negative effect for gastrointestinal disease, and shows no effect for respiratory disease. Since the percentage of illiterate people shows a decreasing trend in time, a negative effect means that gastrointestinal cases have a positive trend in time. We emphasize that the illiteracy might not have a direct relationship with the diseases, its inclusion in the analysis is merely to capture trends. 

\begin{figure}
\begin{center}
\includegraphics[scale=0.45]{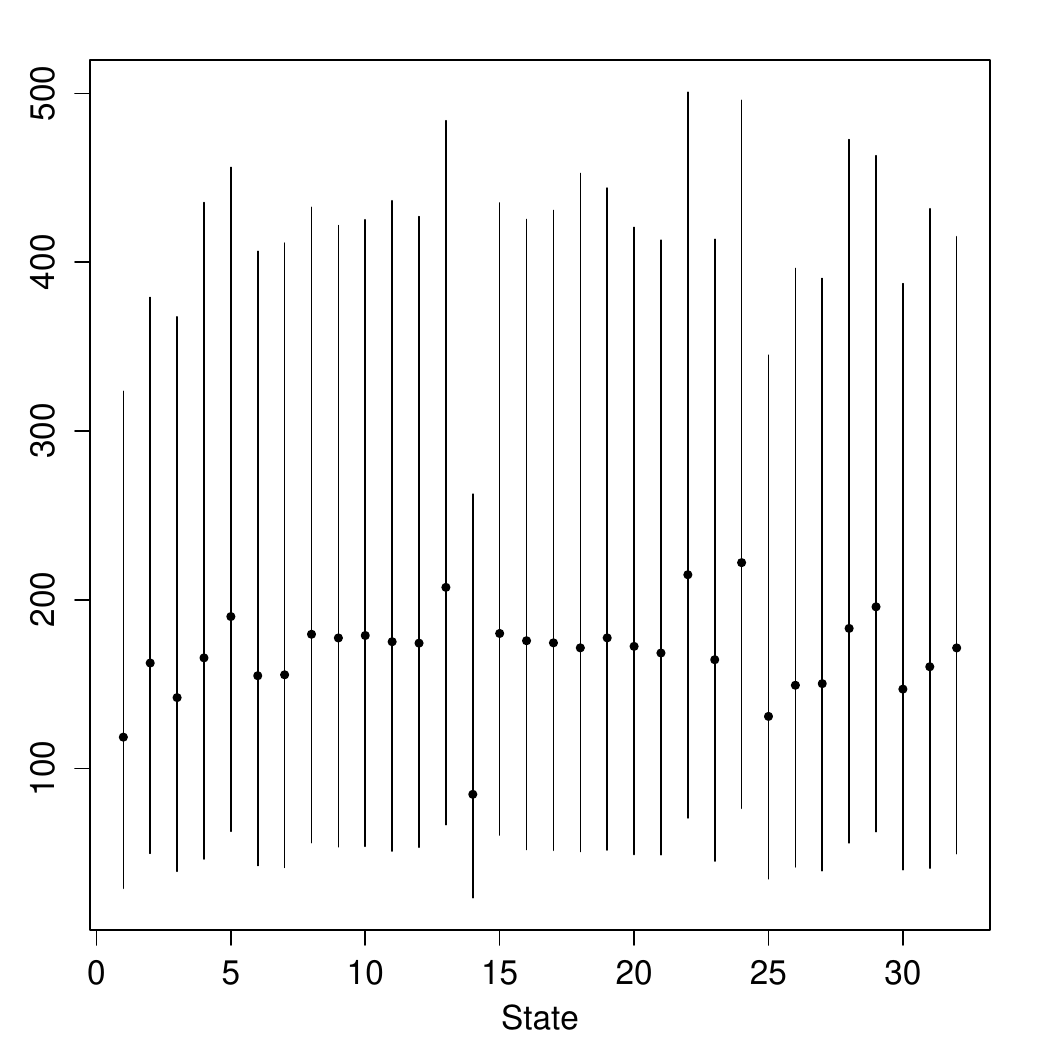}
\includegraphics[scale=0.45]{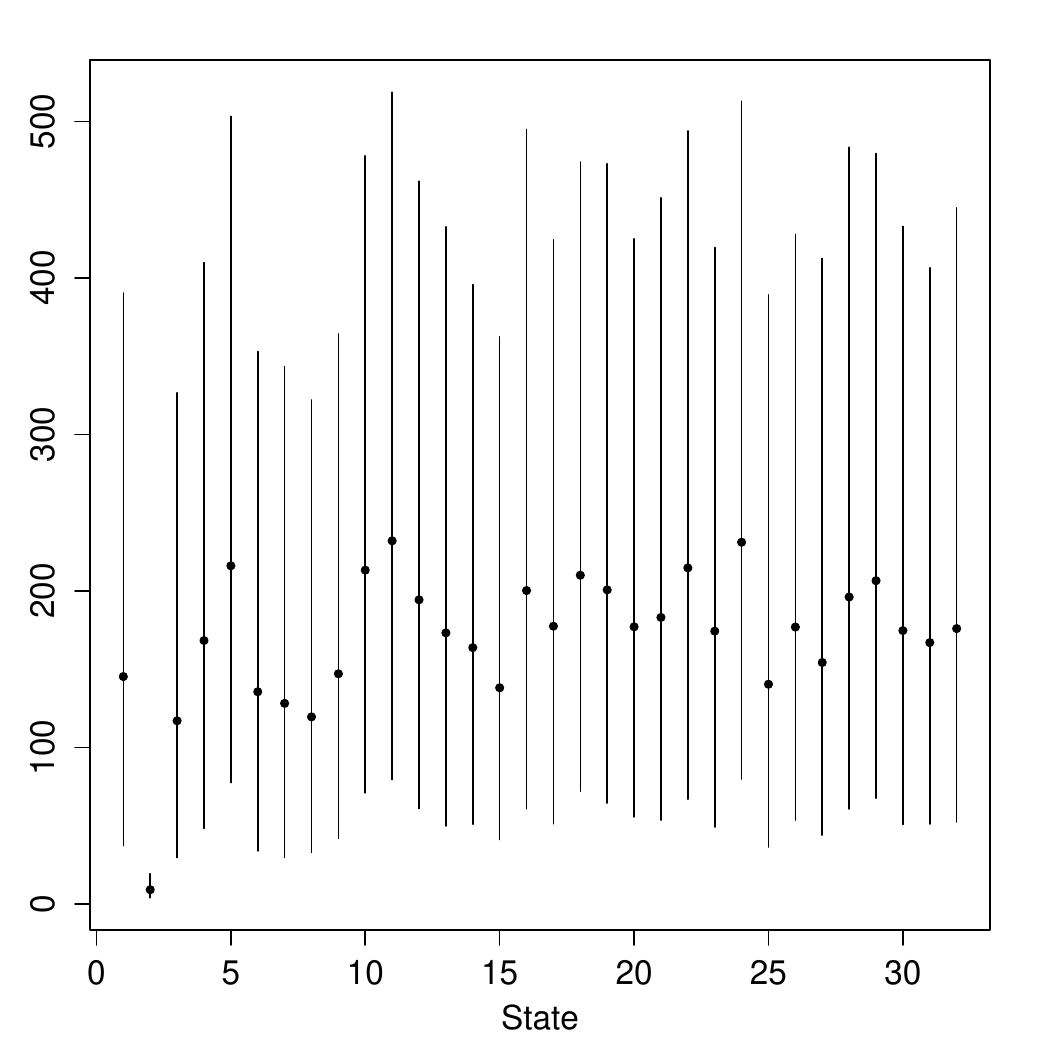}
\caption{\small{Posterior estimates of $\tau_i$, $i=1,\ldots,32$. Point estimates (dot) and 95\% CI (line). Gastrointestinal data (left) and respiratory data (right).}}
\label{fig:tau}
\end{center}
\end{figure}

Interpreting the coefficients for the best models in both datasets, posterior 95\% CI for the intercept are: $\alpha\in(-7.78,-7.42)$ for the gastrointestinal data; and $\alpha\in(-2.19,-1.69)$ for the respiratory data. These values are in accordance with those obtained from model 00. Figure \ref{fig:tau} contains posterior estimates for the precision parameters $\tau_i$, $i=1,\ldots,32$, where the dots correspond to the mean and the vertical lines to 95\% CI. The left panel corresponds to gastrointestinal data and the right panel to respiratory data. In both diseases, point estimates for the precisions lie between 100 and 250 with some exceptions, Jalisco ($i=14$) in the gastrointestinal case, and Baja California ($i=2$) in the respiratory case, whose precisions are lower. There is not particular reason for this to happen, other than the fact that the variability not explained by the model, in these two specific states, is larger than for the other states. 

\begin{figure}
\begin{center}
\includegraphics[scale=0.56]{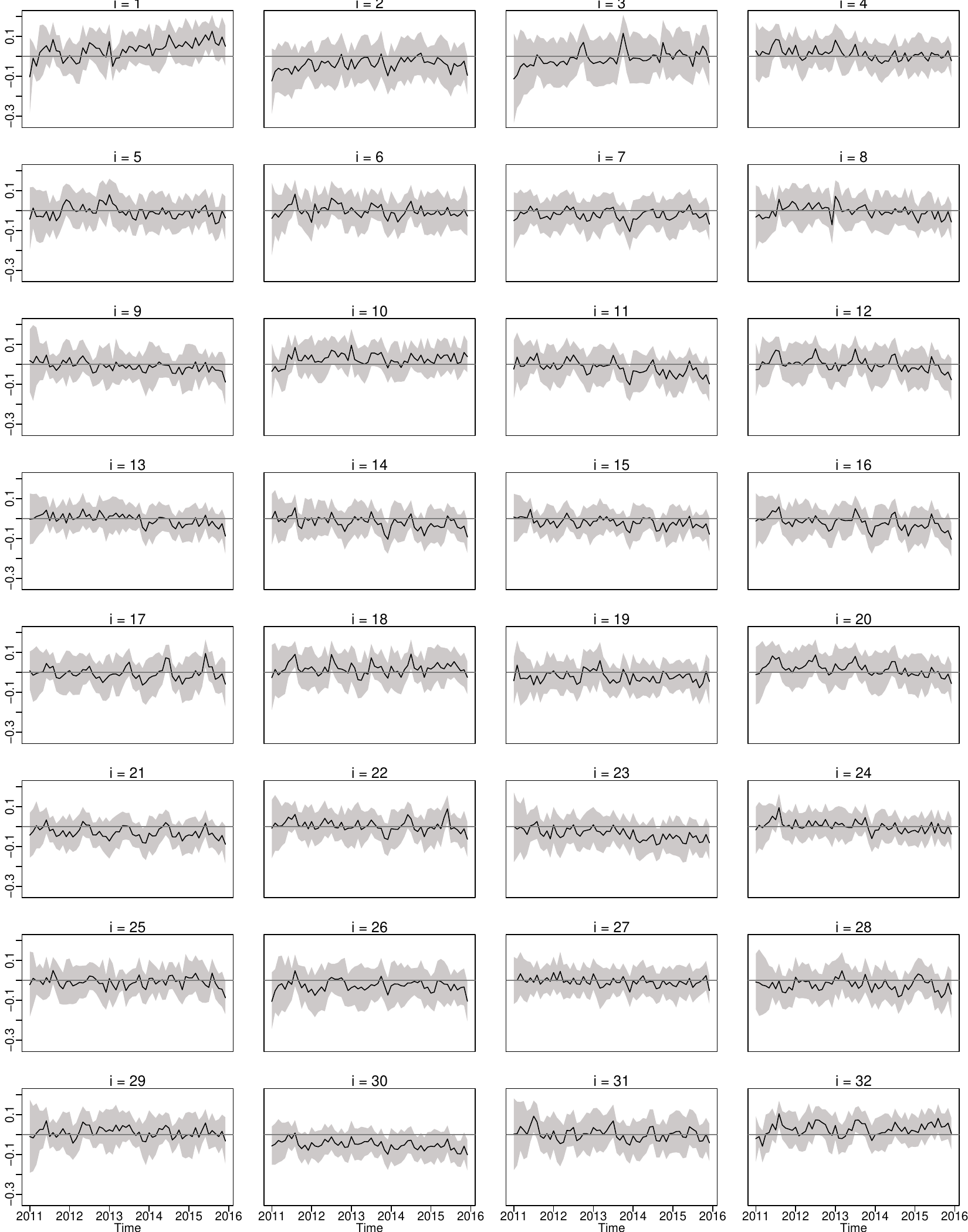}
\caption{\small{Pluvial precipitation effect $\bbeta_1=\{\beta_{i,t,1}\}$, $i=1,\ldots,32$, $t=1,\ldots,60$, for gastrointestinal data. Posterior point estimates (solid line) and 95\% CI (shadows).}}
\label{fig:gbeta1}
\end{center}
\end{figure}

For the gastrointestinal data, posterior estimates of $\bbeta_1$ are shown in Figure \ref{fig:gbeta1}, where we include 32 panels, one for each state, with time series for $t=1,\ldots,60$. Reported are point estimates (solid line) and 95\% CI (shadows). We can see that the estimates are not constant and vary across $i$ and $t$. However, the effect of pluvial precipitation is mainly no significant since most CI's contain the value of zero, with some few exceptions for specific states and specific times. This is the case of Aguascalientes ($i=1$) whose effect shows an increasing trend, but only becomes significant in the rainy season (June, July and August) of 2015, where a 10\% increment in pluvial precipitation produced around 1\% increment in the disease rate. This is in contrast to Veracruz ($i=30$) where a 10\% increment in pluvial precipitation produced around 1\% reduction in the disease rates in the last quarter of the years 2013, 2014 and 2015.

\begin{figure}
\begin{center}
\includegraphics[scale=0.56]{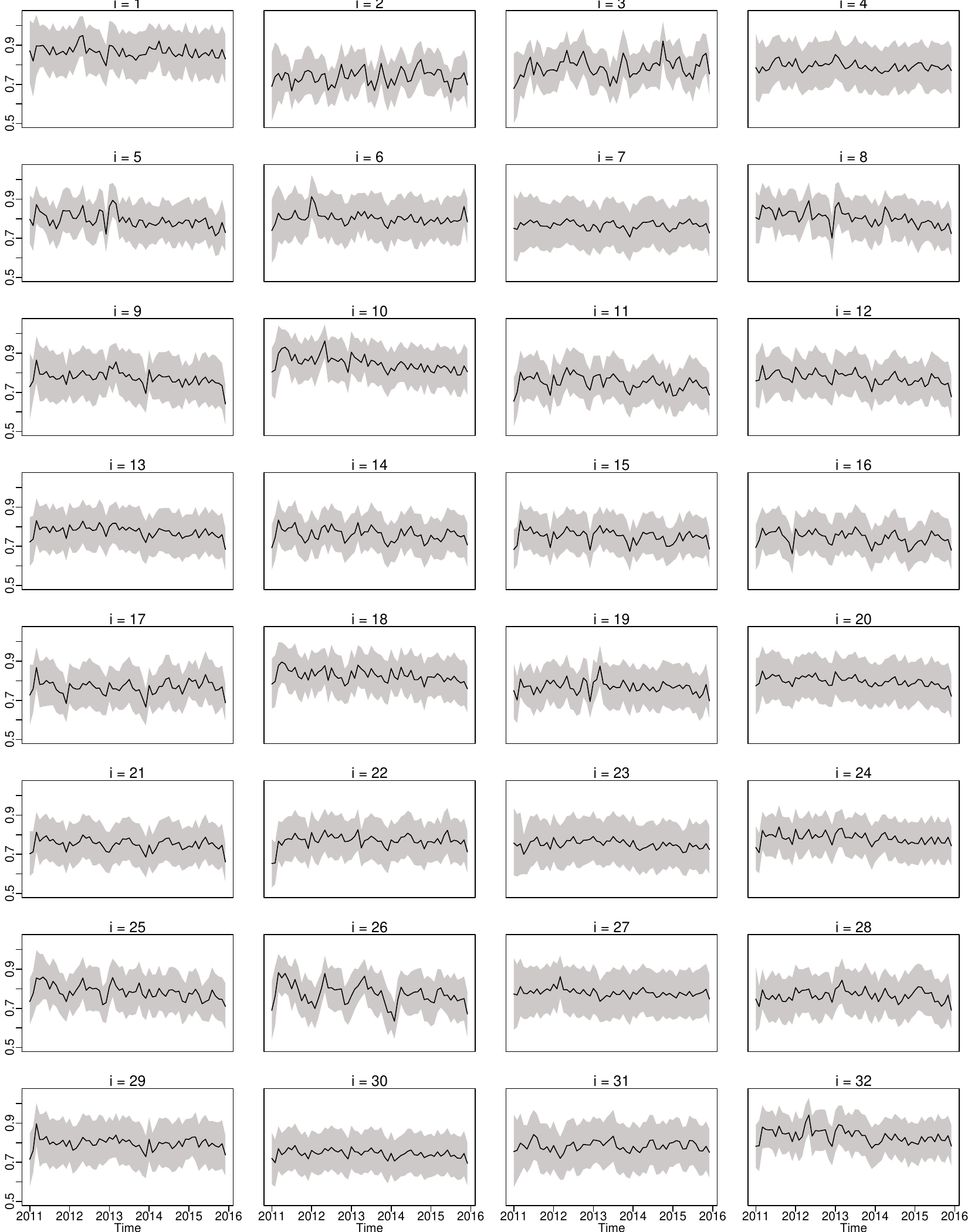}
\caption{\small{Temperature effect $\bbeta_2=\{\beta_{i,t,2}\}$, $i=1,\ldots,32$, $t=1,\ldots,60$, for the gastrointestinal data. Posterior point estimates (solid line) and 95\% CI (shadows).}}
\label{fig:gbeta2}
\end{center}
\end{figure}

On the other hand, Figure \ref{fig:gbeta2} reports posterior estimates of $\bbeta_2$. Again, the estimates are not constant, and for all states and times, posterior CI's only contain positive values, implying a positive significant effect of temperature on the gastrointestinal rates. For all states, posterior means have values between 0.7 and 0.9, which is consistent with the common effect of model 00. For some states, the effect fluctuates more, as is the case of Sonora ($i=26$), and for some others the effect is more steady, as is the case of Tabasco ($i=27$). However, temperature effects are not the same in all states. A 10\% increment in the temperature produced around a 9\% increment in the gastrointestinal rates in Aguscalientes ($i=1$), and around a 7\% increment in Baja California ($i=2$). 

\begin{figure}
\begin{center}
\includegraphics[scale=0.56]{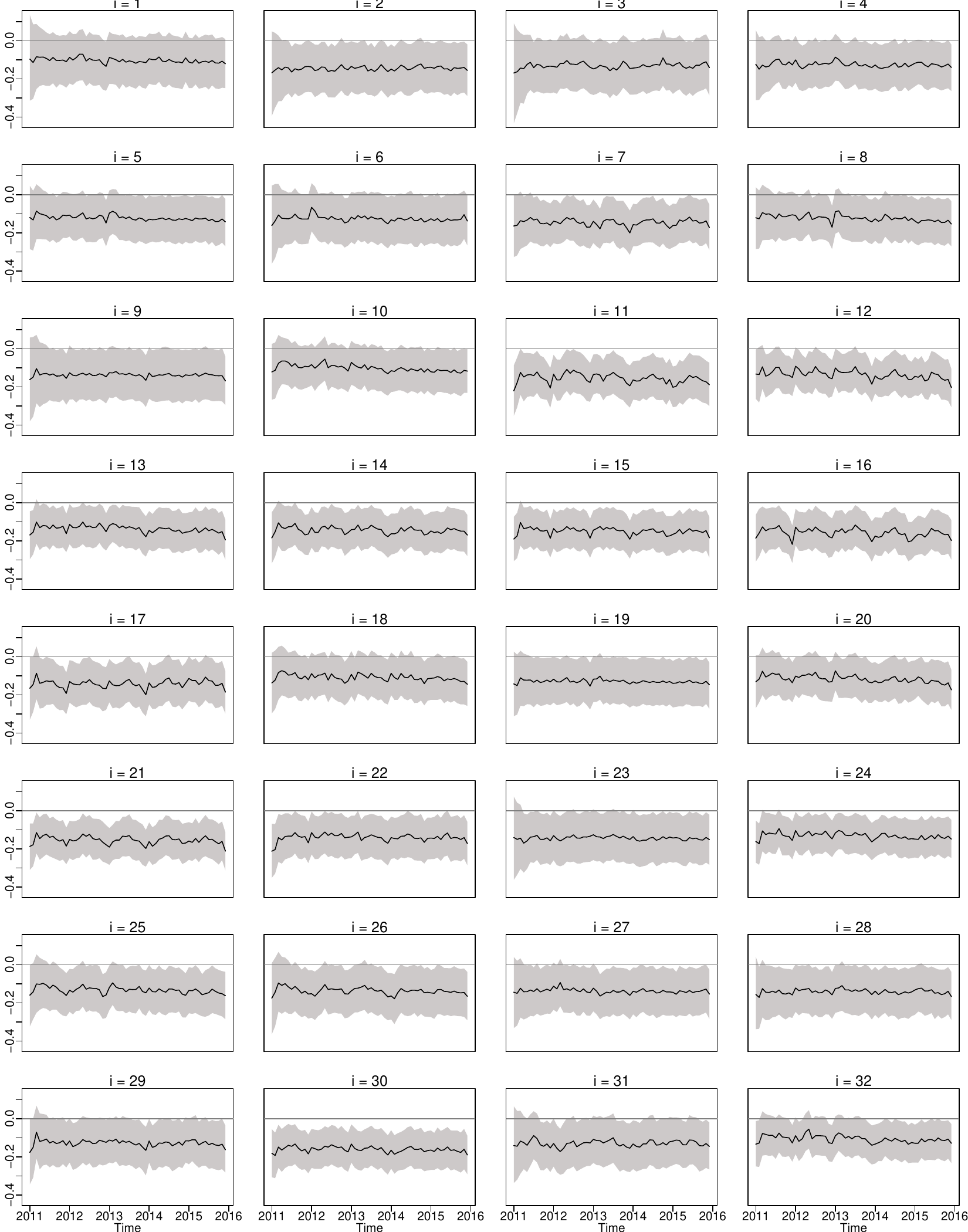}
\caption{\small{Illiteracy effect $\bbeta_3=\{\beta_{i,t,3}\}$, $i=1,\ldots,32$, $t=1,\ldots,60$, for the gastrointestinal data. Posterior point estimates (solid line) and 95\% CI (shadows).}}
\label{fig:gbeta3}
\end{center}
\end{figure}

Finally, Figure \ref{fig:gbeta3} includes posterior estimates of $\bbeta_3$, loosely speaking we can say that the effect of the illiteracy indicator is negative significant for all states, perhaps for some states the credibility should be 90\% instead of 95\%. The point estimates lie between $-0.1$ and $-0.2$, which is also consistent with the common effect of model 00. This implies that there is a slight positive trend in the gastrointestinal rates in all states. 

\begin{figure}
\begin{center}
\includegraphics[scale=0.56]{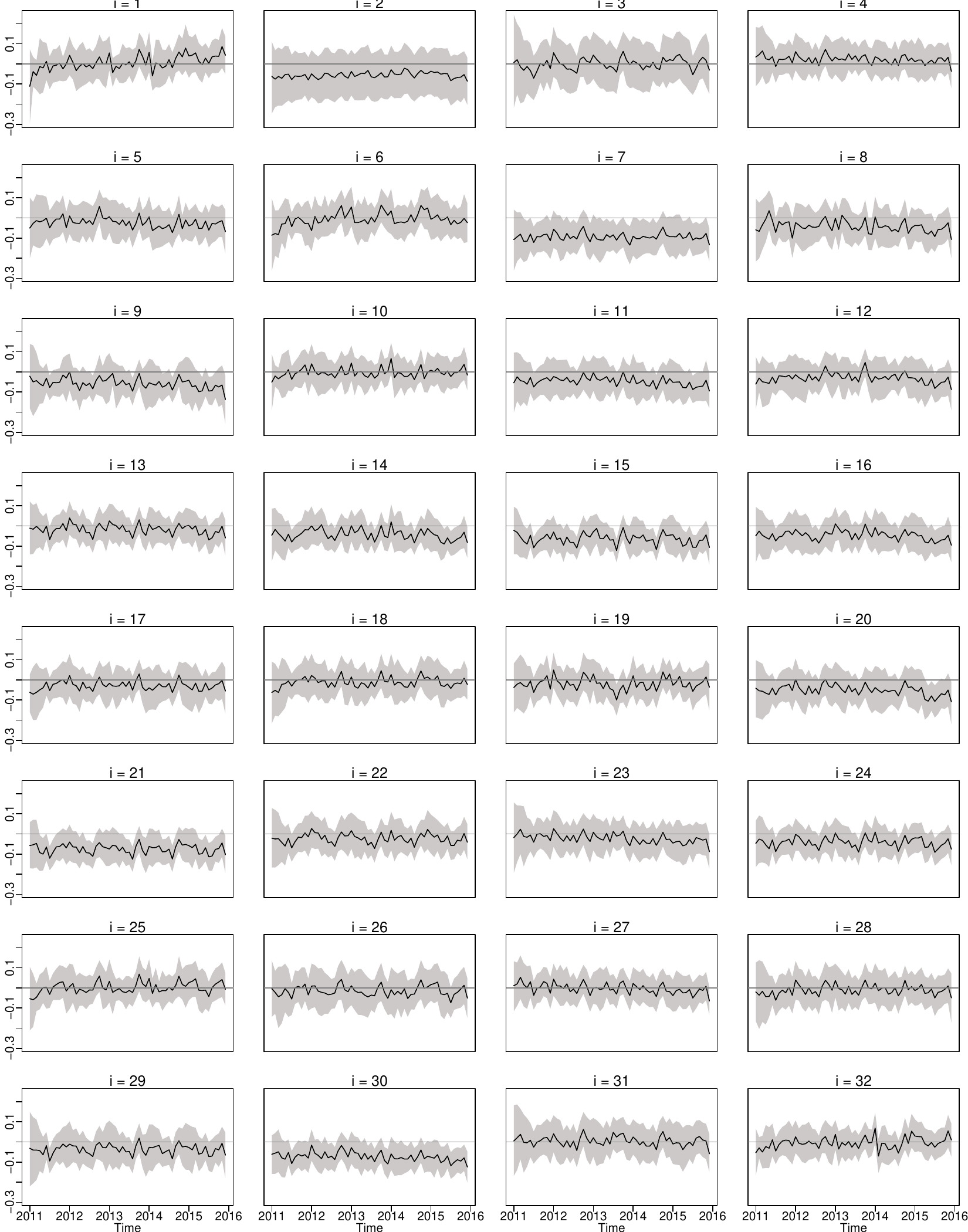}
\caption{\small{Pluvial precipitation effect  $\bbeta_1=\{\beta_{i,t,1}\}$, $i=1,\ldots,32$, $t=1,\ldots,60$, for the respiratory data. Posterior point estimates (solid line) and 95\% CI (shadows).}}
\label{fig:rbeta1}
\end{center}
\end{figure}

For the respiratory data, posterior estimates of $\bbeta_1$ are included in Figure \ref{fig:rbeta1}. There are only three states with a negative significant effect of pluvial precipitation for all months, these are Chiapas ($i=7$), Puebla ($i=21$) and Veracruz ($i=30$), where a 10\% increment in the pluvial precipitation produced a 1\% decrement in the disease rates. For other states like the State of Mexico ($i=15$) and San Luis Potosi ($i=24$), only the rain season (June, July and August) shows a negative significant effect of the same amount as in the former three states. For the rest of the states there is no effect. 

\begin{figure}
\begin{center}
\includegraphics[scale=0.56]{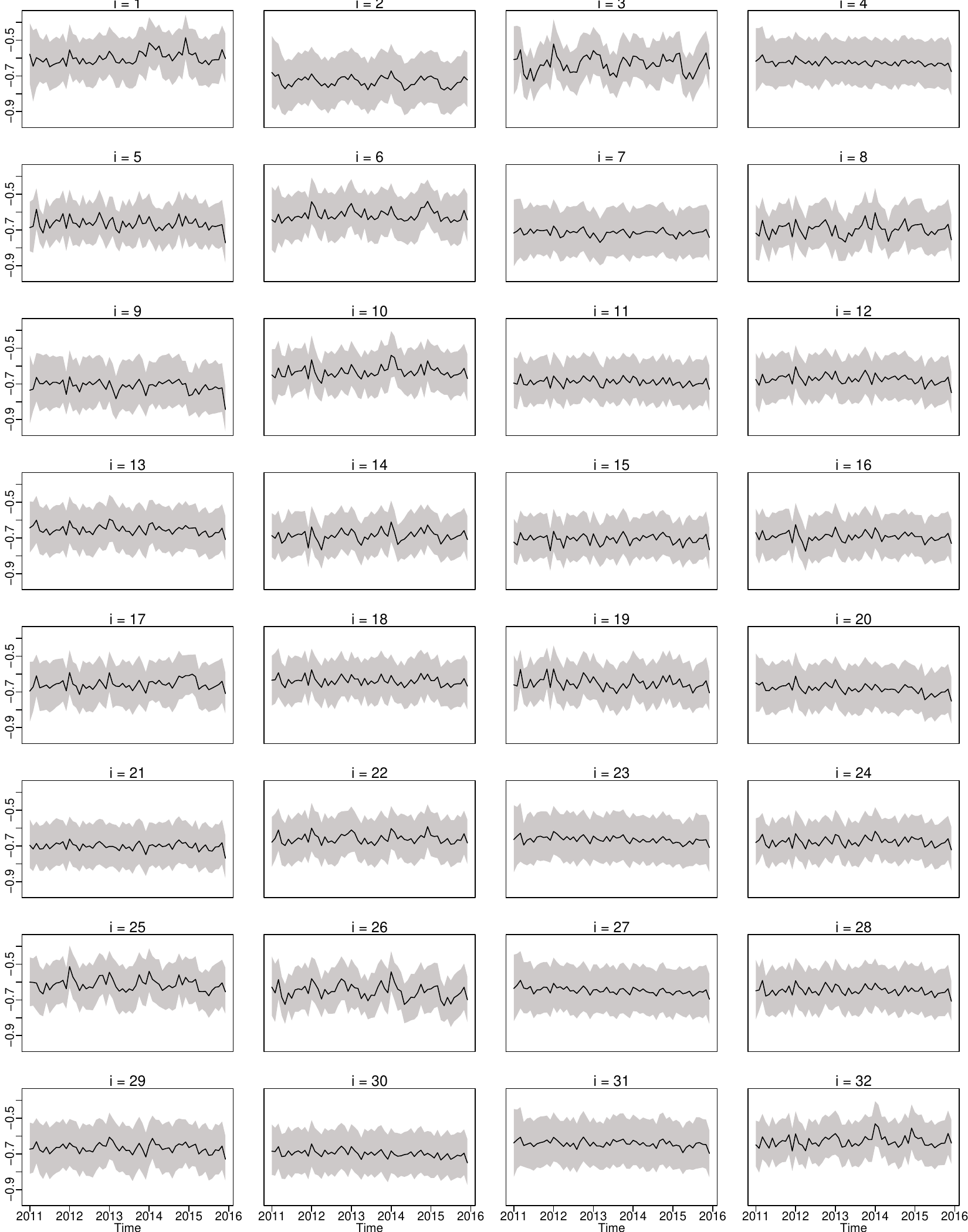}
\caption{\small{Temperature effect $\bbeta_2=\{\beta_{i,t,2}\}$, $i=1,\ldots,32$, $t=1,\ldots,60$, for the respiratory data. Posterior point estimates (solid line) and 95\% CI (shadows).}}
\label{fig:rbeta2}
\end{center}
\end{figure}

Posterior estimates of $\bbeta_2$ are shown in Figure \ref{fig:rbeta2}. It is clear that the temperature has a negative significant effect for all states and all months in the respiratory rates. Posterior means lie between $-0.6$ and $-0.7$, consistent with the common effect estimate of model 00. In particular, a 10\% increment in the temperature produced around 6\% decrement in Aguascalientes ($i=1$) and around a 7.5\% decrement in Baja California ($i=2$), in the respiratory rates. 

\begin{figure}
\begin{center}
\includegraphics[scale=0.56]{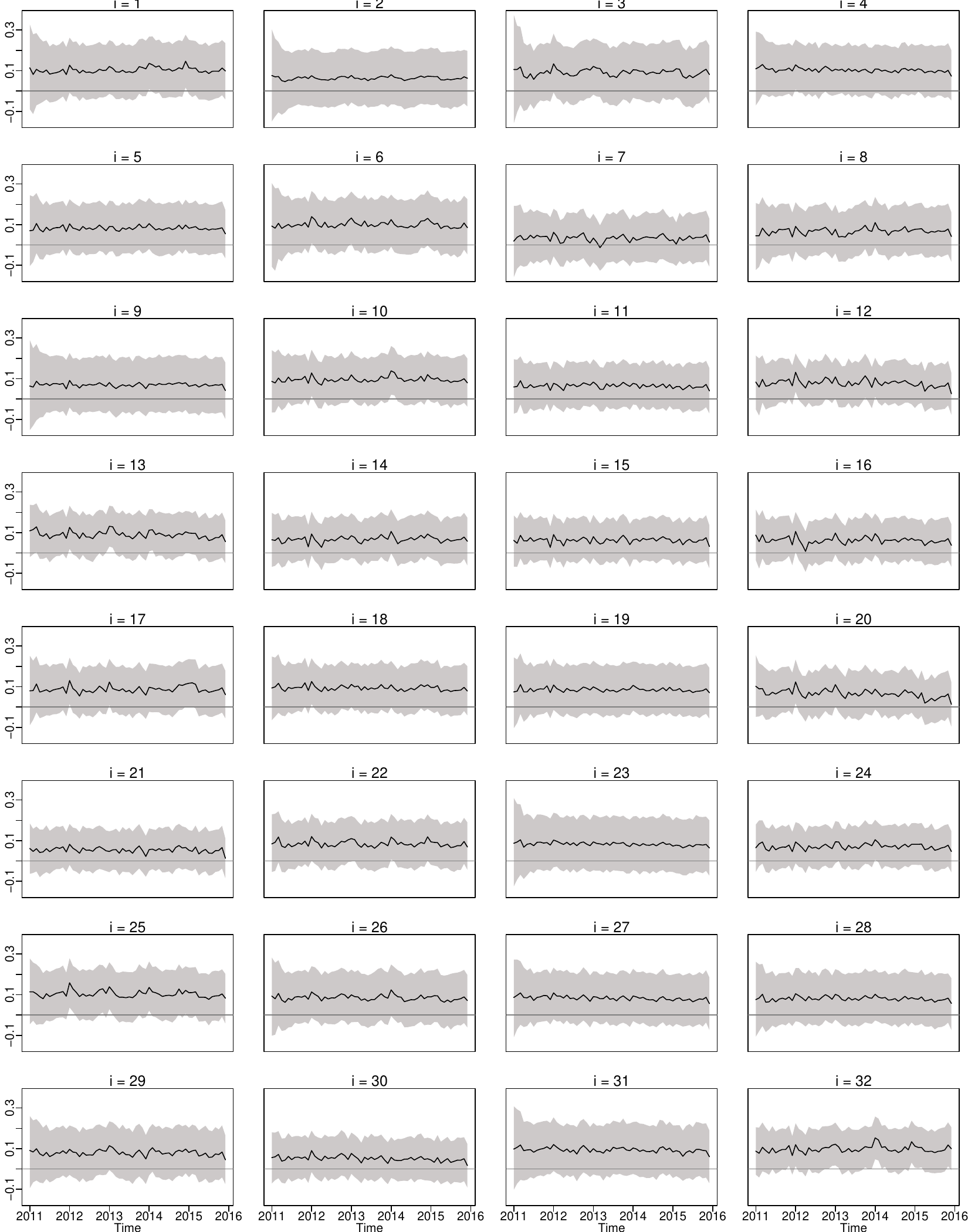}
\caption{\small{Illiteracy effect $\bbeta_3=\{\beta_{i,t,3}\}$, $i=1,\ldots,32$, $t=1,\ldots,60$, for the respiratory data. Posterior point estimates (solid line) and 95\% CI (shadows).}}
\label{fig:rbeta3}
\end{center}
\end{figure}

Finally, posterior estimates of $\bbeta_3$ are given in Figure \ref{fig:rbeta3}. For most states there is no significant effect of the illiteracy indicator in the respiratory rates, perhaps the exceptions are Hidalgo ($i=13$), Sinaloa ($i=25$) and Zacatecas ($i=32$) which show a positive significant effect, implying that there is a negative trend in the respiratory rates in these states.

\section{Concluding remarks}
\label{sec:conclusion}

We have proposed a space-time dependent process, STN, which has a multivariate normal distribution and identically distributed marginal distributions. We use our process as a prior distribution for the coefficients in a regression model. The prior is very flexible so it allows us to identify areas and times where explanatory variables show differentiated effects, as was the case in the study of gastrointestinal and respiratory diseases in Mexico.  

When comparing our model with competitors previously proposed in the literature, our model outperforms all of them. We believe that none of the typical priors (CAR, AR and MCAR), borrow enough strength from spatial and temporal neighbours. The reason is that there is a single data point ($(y_{i,t},x_{i,t})$) to update the coefficients $\alpha$ and $\beta_{i,t,k}$, $k=1,\ldots,p$. A more structured prior as the one we are proposing is preferred in this varying coefficients regression setting.

Other use of our space-time process is to model responses with spatial and temporal dependence. In this case the model would be used as a sampling distribution (likelihood) instead of a prior. Additionally, to make our model even more flexible, instead of setting a value for each $c_{i,t}$ as we did here, we could put a hyper-prior (hierarchical) distribution on all of them so the data can help us determine their best value. 

An alternative construction to define a spatial process that gives a different strength to each connection is the following. Instead of defining a latent parameter for each location $i$, we define a latent parameter for each connection. That is, if location $i$ is neighbour of location $j$, in notation $i\sim j$, then we define a latent $\gamma_{j}^{i}$. Of course we do require a symmetry condition such that $\gamma_{j}^{i}\equiv\gamma_{i}^{j}$. If we now let $\bgamma=\{\gamma_{j}^{i}\}$, a different model for the distribution of spatial $\bbeta=\{\beta_i\}$ would be
\begin{eqnarray}
\nonumber
\beta_{i}\mid\bgamma&\simind&\no\left(\frac{c_0m_0+\sum_{j\in\partial_{i}}c_{j}^{i}\gamma_{j}^{i}}{c_0+\sum_{j\in\partial_{i}}c_{j}^{i}},\,c_0+\sum_{j\in\partial_{i}}c_{j}^{i}\right)\\
\label{eq:stnormal2}
\gamma^{i}_{j}\mid\omega&\simind&\no(\omega,c^{i}_{j}) \\
\nonumber
\omega&\sim&\no(m_0,c_0)
\end{eqnarray}
where $m_0\in\Ree$, $c_0>0$ and $c_{i}^{j}>0$ if $i\sim j$, and $c_{i}^{j}=0$ otherwise, similarly $\gamma_{i}^{j}=0$ with probability one if $i\not\sim j$, for $i,j=1,\ldots,n$, and $\partial_i$ a set of spatial neighbours, as that given in (s) at the beginning of Section  \ref{sec:model}. Now, parameters $\bc=\{c_{i}^{j}\}$ determine the strength of dependence between locations $i$ and $j$. It can be proven that the joint distribution of $\bbeta$ is a multivariate normal with mean $m_0\bone_n$ and correlation between any two $\beta_{i}$ and $\beta_{j}$ for $i\neq j$ given by
$$\Cr(\beta_{i},\beta_{j})=\frac{c_0c_{j}^{i}I(i\sim j)+ \left(\sum_{\partial_{i}}c_{k}^{i}\right)\left(\sum_{\partial_{j}}c_{k}^{j}\right)} {\left(c_0+\sum_{\partial_{i}}c_{k}^{i}\right)\left(c_0+\sum_{\partial_{j}}c_{k}^{j}\right)},$$
which clarifies that a larger value of $c_{j}^{i}$ will induce a larger association between neighbours $\beta_{i}$ and $\beta_{j}$. Although the pairwise dependence is more flexible due to the existence of a linking parameter $c_{j}^{i}$, the correlation between two regions that are second order neighbours (i.e., are not neighbours but share a common neighbour) will be a lot less than that induce by the first model \eqref{eq:stnormal1}. Moreover, extending this spatial process to a spatio-temporal process is not trivial due to the lack of symmetry in the temporal dependence.

\section*{Acknowledgements}
This research was supported by \textit{Asociaci\'on Mexicana de Cultura, A.C.}--Mexico. The author is grateful to the United Nations Development Programme in Mexico and to two anonymous referees for their insightful comments.

\bibliographystyle{natbib}

\begin{thebibliography}{99}

\bibitem[Banerjee et al., 2003]{banerjee&al:03}
Banerjee, S., Wall, M. M. and Carlin, B. P. (2003). Frailty modeling for spatially correlated survival data, with application to infant mortality in Minnesota. {\it Biostatistics} {\bf 4}, 123--142.

\bibitem[Besag et al., 1991]{besag&al:91}
Besag, J., York, J. and Molli\'e, A. (1991). Bayesian image restoration, with two applications in spatial statistics (with discussion). {\it Annals of the Institute of Statistical Mathematics} {\bf 43}, 1--59.

\bibitem[Best et al., 2005]{best&al:05}
Best, N., Richardson, S. and Thomson, A. (2005). A Comparison of Bayesian spatial models for disease mapping. {\it Statistical Methods in Medical Research} {\bf 14}, 35--59.

\bibitem[Cai et al., 2013]{cai&al:13}
Cai, B., Lawson, A.B., Hossain, M., Choi, J. Kirby, R.S. and Liu, J. (2013). Bayesian semiparametric model with spatially–temporally varying
coefficients selection. {\it Statistics in Medicine} {\bf 32}, 3670--3685.

\bibitem[Choi et al., 2012]{choi&al:12}
Choi, J. Lawson, A.B., Cai, B., Hossain, M., Kirby, R.S. and Liu, J. (2012). A Bayesian latent model with spatio-temporally varying coefficients in low birth weight incidence data. {\it Statistical Methods in Medical Research} {\bf 21}, 445--456.

\bibitem[D'Amato et al., 2014]{damato&al:14}
D’Amato, G., Cecchi, L., D’Amato, M., Annesi-Maesano, I. (2014). Climate change and respiratory diseases. {\it European Respiratory Review} {\bf 23}, 161--169.

\bibitem[Geisser and Eddy, 1979]{geisser&eddy:79}
Geisser, S. and Eddy, W.F. (1979). A predictive approach to model selection. {\it Journal of the American Statistical Association} {\bf 74}, 153--160.

\bibitem[Gelfand et al., 2003]{gelfand&al:03}
Gelfand, A.E., Hyon-Jung, K., Sirmans, C.F. and Banerjee, S. (2003). Spatial modeling with spatially varying coefficients processes. {\it Journal of the American Statistical Association} {\bf 98}, 387--396.

\bibitem[Gelfand and Vounatsou, 2003]{gelfand&vounatsou:03}
Gelfand, A.E. and Vounatsou, P. (2003). Proper multivariate conditional autoregressive models for spatial data analysis. {\it Biostatistics} {\bf 4}, 11--25.

\bibitem[Goicoa et al., 2018]{goicoa&al:18}
Goicoa, T., Adin, A. and Ugarte, M.D. (2018). In spatio-temporal disease mapping models, identifiability constraints affect PQL and INLA results.  {\it Stochastic Environmental Research and Risk Assessment} {\bf 32}, 749--770.

\bibitem[Jara et al., 2013]{jara&al:13}
Jara, A., Nieto-Barajas, L.E. and Quintana, F. (2013). A time series model for responses on the unit interval. {\it Bayesian Analysis} {\bf 8}, 723--740. 

\bibitem[Jin and Carlin, 2005]{jin&carlin:05}
Jin, X. and Carlin , B.P. (2005). Multivariate parametric spatiotemporal models for county level breast cancer survival data. {\it Lifetime Data Analysis} {\bf 11}, 5--27. 

\bibitem[Lawson, 2009]{lawson:09}
Lawson, A. B. (2009). {\it Bayesian Disease Mapping: Hierarchical Models in Spatial Epidemiology}. Boca Raton, FL: Chapman and Hall/CRC.

\bibitem[MacNab and Dean, 2001]{macnab&dean:01} 
MacNab, Y.C. and Dean, C.B. (2001). Autoregressive spatial smoothing and temporal spline smoothing for mapping rates. {\it Biometrics} {\bf 57}, 949--956.

\bibitem[McCullagh and Nelder, 1989]{mccullagh&nelder:89} 
McCullagh, P. and Nelder, J.A. (1989). {\it Generalized linear models}. Chapman and Hall, London. 

\bibitem[Morral-Puigmala, 2018]{morral&al:18}
Morral-Puigmala, C., Martínez-Solanasa, E., Villanueva, C.M. and 
Basagaña X. (2018). Weather and gastrointestinal disease in Spain: A retrospective time series regression study. {\it Environment International} {\bf 121}, 649--657.

\bibitem[Nieto-Barajas and Huerta, 2017]{nieto&huerta:17} 
Nieto-Barajas, L.E. and Huerta, G. (2017). Spatio-temporal pareto modelling of heavy-tail data. {\it Spatial Statistics} {\bf 20}, 92--109. 

\bibitem[Nieto-Barajas and Bandyopadhyay, 2013]{nieto&bandyopadhyay:13} 
Nieto-Barajas, L.E. and Bandyopadhyay, D. (2013). A zero-inflated spatial gamma process model with applications to disease mapping. {\it Journal of Agricultural, Biological and Environmental Statistics} {\bf 18}, 137--158. 

\bibitem[Schr\"odle and Held, 2010]{schrodle&held:10}
Schr\"odle, B. and Held, L. (2010). Spatio-temporal disease mapping using INLA. {\it Environmetrics} {\bf 22}, 725--734.

\bibitem[Smith and Roberts, 1993]{smith&roberts:93}
Smith, A. and Roberts, G. (1993). Bayesian computations via the Gibbs sampler and related Markov chain Monte Carlo methods. {\it Journal of the Royal Statistical Society, Series B} {\bf 55}, 3--23.

\bibitem[Spiegelhalter et al., 2002]{spiegelhalter&al:02}
Spiegelhalter, D.J., Best, N.G., Carlin, B.P. and van der Linde, A. (2002). Bayesian measures of model complexity and fit (with discussion). {\it Journal of the Royal Statistical Society, Series B} {\bf 64}, 583--639.

\bibitem[Torabi and Rosychuk, 2010]{torabi&rosychuk:10}
Torabi, M. and Rosychuk, M.J. (2010). Spatio-temporal modelling of disease mapping of rates. {\it The Canadian Journal of Statistics} {\bf 38}, 698--715.

\bibitem[Tierney, 1994]{tierney:94}
Tierney, L. (1994). Markov chains for exploring posterior distributions. {\it Annals of Statistics} {\bf 22}, 1701-1762.

\bibitem[Waller et al., 1997]{waller&al:97}
Waller, L.A., Carlin, B.P., Xia, H. and Gelfand, A.E. (1997). Hierarchical spatio-temporal mapping of disease rates. {\it Journal of the American Statistical Association} {\bf 92}, 607--617.

\end{thebibliography}

\newpage

\begin{table}
\caption{Simulated data: Goodness of fit measures for different models defined by neighbourhood type and temporal dependence $q$. Bold numbers correspond to the best fit.}
\label{tab:sim}{\normalsize
\begin{center}
\begin{tabular}{cc|cc|cc} \hline \hline
 & & \multicolumn{2}{c|}{Scenario 1} & \multicolumn{2}{c}{Scenario 2} \\
Type & $q$ & LPML & DIC & LPML & DIC \\ \hline
$s$	&	$-$	&	$-42.77$	&	84.02	&	$-47.38$	&	92.27	\\
$t_q$	&	1	&	$-43.82$	&	83.23	&	$-47.52$	&	92.55	\\
$t_q$	&	2	&	$\bf -40.43$	&	{\bf 78.48}	&	$-48.40$	&	93.69	\\
$s+t_q$	&	1	&	$-43.65$	&	81.93	&	$\bf -46.70$	&	{\bf 90.58}	\\
$s+t_q$	&	2	&	$-43.76$	&	81.60	&	$-48.35$	&	95.06	\\
$s\times t_q$	&	1	&	$-43.10$	&	82.17	&	$-47.13$	&	91.79	\\
$s\times t_q$	&	2	&	$-44.86$	&	85.54	&	$-47.81$	&	93.99	\\
\hline \hline
\end{tabular}
\end{center}}
\end{table}

\begin{table}
\caption{Mexican states and adjacencies.}
\label{tab:adj}{\footnotesize
\begin{center}
\begin{tabular}{cll|cll} \hline \hline
ID & State & Adjacent states & ID & State & Adjacent states \\ \hline
1 & Aguascalientes & 14,32 & 17 & Morelos & 9,12,15,21 \\
2 & Baja California & 3,26 & 18 & Nayarit & 10,14,25,32 \\
3 & Baja California Sur & 2 & 19 & Nuevo Leon & 5,24,28,32 \\
4 & Campeche & 23,27,31 & 20 & Oaxaca & 7,12,21,30 \\
5 & Coahuila & 8,10,19,32 & 21 & Puebla & 12,13,15,17,20,29,30 \\
6 & Colima & 14,16 & 22 & Queretaro & 11,13,15,16,24 \\
7 & Chiapas & 20,27,30 & 23 & Quintana Roo & 4,31 \\
8 & Chihuahua & 5,10,25,26 & 24 & San Luis Potosi & 11,13,14,19,22,28,30,32 \\
9 & Distrito Federal & 15,17 & 25 & Sinaloa & 8,10,18,26 \\
10 & Durango & 5,8,18,25,32 & 26 & Sonora & 2,8,25 \\
11 & Guanajuato & 14,16,22,24,32 & 27 & Tabasco & 4,7,30 \\
12 & Guerrero & 15,16,17,20,21 & 28 & Tamaulipas & 19,24,30 \\
13 & Hidalgo & 15,21,22,24,29,30 & 29 & Tlaxcala & 13,15,21 \\
14 & Jalisco & 1,6,11,16,18,24,32 & 30 & Veracruz & 7,13,20,21,24,27,28 \\
15 & Mexico & 9,12,13,16,17,21,22,29 & 31 & Yucatan & 4,23 \\
16 & Michoacan & 6,11,12,14,15,22 & 32 & Zacatecas & 1,5,10,11,14,18,19,24 \\
\hline \hline
\end{tabular}
\end{center}}
\end{table}

\begin{table}
\caption{Real data: Goodness of fit measures for different models defined by neighbourhood type, temporal dependence $q$ and association parameter $c_{i,t}$. The median neighbourhood size and competing models fittings are also reported. Gastrointestinal data (gastro) and respiratory data (resp). Bold numbers correspond to the best fit in each neighbourhood definition.}
\label{tab:gof}
\begin{center}
\begin{tabular}{cccc|cc|cc} \hline \hline
 & & & & \multicolumn{2}{c|}{gastro} & \multicolumn{2}{c}{resp} \\
Type & $q$ & $\mbox{med}|\partial_{i,t}|$ & $c_{i,t}$ & LPML & DIC &LPML & DIC \\ \hline 
$s$ & -- & 5 & 10 & $-88$ & $-85$ & $-95$ & $17$ \\
$s$ & -- & 5 & 50 & $\bf 549$ & $\bf -1332$ & $\bf 497$ & $\bf -1140$ \\
$s$ & -- & 5 & 100 & $514$ & $-1304$ & $442$ & $-1102$ \\ \hline
$t_q$ & 3 & 4 & 10 & $-202$ & $257$ & $-161$ & $184$ \\
$t_q$ & 6 & 7 & 10 & $172$ & $-432$ & $188$ & $-461$ \\
$t_q$ & 3 & 4 & 50 & $634$ & $-1399$ & $569$ & $-1308$ \\
$t_q$ & 6 & 7 & 50 & $744$ & $-1716$ & $\bf 769$ & $\bf -1770$ \\
$t_q$ & 3 & 4 & 100 & $\bf 776$ & $\bf -1748$ & $678$ & $-1627$ \\
$t_q$ & 6 & 7 & 100 & $673$ & $-1668$ & $689$ & $-1715$ \\ \hline
$s+t_q$ & 3 & 8 & 10 & $208$ & $-552$ & $170$ & $-492$ \\
$s+t_q$ & 6 & 11 & 10 & $370$ & $-883$ & $394$ & $-954$ \\
$s+t_q$ & 3 & 8 & 50 & $810$ & $-1867$ & $738$ & $-1703$ \\
$s+t_q$ & 6 & 11 & 50 & $\bf 974$ & $\bf -2199$ & $\bf 982$ & $\bf -2148$ \\
$s+t_q$ & 3 & 8 & 100 & $740$ & $-1757$ & $690$ & $-1587$ \\
$s+t_q$ & 6 & 11 & 100 & $817$ & $-1955$ & $935$ & $-2061$ \\ \hline
$s\times t_q$ & 3 & 20 & 10 & $662$ & $-1490$ & $581$ & $-1342$ \\
$s\times t_q$ & 6 & 35 & 10 & $\bf 734$ & $\bf -1740$ & $\bf 734$ & $\bf -1665$ \\
$s\times t_q$ & 3 & 20 & 50 & $504$ & $-1304$ & $379$ & $-1043$ \\
$s\times t_q$ & 6 & 35 & 50 & $224$ & $-631$ & $285$ & $-752$ \\
$s\times t_q$ & 3 & 20 & 100 & $205$ & $-595$ & $178$ & $-551$ \\
$s\times t_q$ & 6 & 35 & 100 & $1$ & $-88$ & $66$ & $-206$ \\ \hline
\multicolumn{4}{c|}{CAR, $\rho=0.99$, $\tau_c=0.1$} & $-2652$ & $6042$ & $-2620$ & $5967$ \\
\multicolumn{4}{c|}{AR, $\varphi=0.99$, $\tau_{ar}=0.1$} & $-799$ & $1627$ & $-806$ & $1803$ \\
\multicolumn{4}{c|}{MCAR, $\rho=\varphi=0.99$, $\tau_{ar}=0.1$} & $-97$ & $127$ & $-96$ & $102$ \\ \hline
\multicolumn{4}{c|}{Model 0} & $211$ & $-417$ & $513$ & $-1021$ \\
\multicolumn{4}{c|}{Model 00} & $-408$ & $819$ & $-485$ & $970$ \\
\hline \hline
\end{tabular}
\end{center}
\end{table}

\newpage

\end{document}